\documentclass{article}
\usepackage{graphicx} 
\usepackage[a4paper, total={6in, 8in}]{geometry}
\usepackage{apacite}
\usepackage{amssymb}
\usepackage{amsmath}
\usepackage{mathtools}
\usepackage{trimclip}
\usepackage{caption}
\usepackage{subcaption}
\usepackage{authblk}

\newcommand\picdims[4][]{%
  \setbox0=\hbox{\includegraphics[#1]{#4}}%
  \clipbox{.5\dimexpr\wd0-#2\relax{} %
           .5\dimexpr\ht0-#3\relax{} %
           .5\dimexpr\wd0-#2\relax{} %
           .5\dimexpr\ht0-#3\relax}{\includegraphics[#1]{#4}}}

\newcommand\xqed[1]{%
  \leavevmode\unskip\penalty9999 \hbox{}\nobreak\hfill
  \quad\hbox{#1}}
\newcommand\demo{\xqed{$\triangle$}}

\title{A geometrical perspective on parametric psychometric models}
\author{Francis Tuerlinckx\textsuperscript{1}}
\affil{University of Leuven, Leuven, Belgium}
\date{\today}

\begin{document}

\maketitle

\begin{abstract}
Psychometrics and quantitative psychology rely strongly on statistical models to measure psychological processes. As a branch of mathematics, geometry is inherently connected to measurement and focuses on properties such as distance and volume. However, despite the common root of measurement, geometry is currently not used a lot in psychological measurement. In this paper, my aim is to illustrate how ideas from non-Euclidean geometry may be relevant for psychometrics.
\end{abstract}

\section{Introduction}

\footnotetext[1]{ORCID: \url{http://orcid.org/0000-0002-1775-7654}. This paper is the result of follow-up work based on my presidential address given at IMPS 2019 in Santiago de Chile. Thanks to Wolf and Geert Vanpaemel for the information on the etymology of the word parameter. Many thanks to Joeri Van der Veken for proofreading the paper (although all remaining errors are the responsability of the author). The code for generating the figures in this paper and performing the calculations can be found on \url{https://osf.io/awz94/}. Contact information: francis.tuerlinckx@kuleuven.be.}

The term "geometry" derives from the Ancient Greek words for earth and measurement. Although it is deeply rooted in measurement and science, geometry is largely absent from current psychological measurement, quantitative or mathematical psychology (although there are exceptions, \citeNP{dzhafarov1999fechnerian}). Instead, researchers and practitioners in these fields rely heavily on statistical models to measure and understand behavior and geometry is rarely considered, except for illustrative purposes.

Classical or Euclidean geometry starts from objects such as points and lines, which are situated in a flat space and for which features such as distances, areas, and angles are studied. Non-Euclidan geometry studies curved spaces. A shared key aspect of both Euclidean and non-Euclidean geometry is that the results should be independent from a particular choice of coordinate system. The distance between two points in the Euclidean space is does not depend on which coordinate system is used.  

Although rarely done so, the statistical models used in psychometrics can also be studied from a geometrical perspective. This will be done in this paper. Thus, I will explore an alternative way of looking at statistical models, and thereby focusing on properties that are invariant with respect to a chosen coordinate system. In a statistical context, this means that properties will be studied that remain the same regardless of the chosen parametrization. The tools needed for such a geometrical inquiry of psychometrical models are based on non-Euclidean geometry. 

\section{Preliminaries}

Assume that we are studying a phenomenon that can be quantified with a (scalar) random variable $y$. For example, this may be the number of correct responses on small test of $m$ items, in which case $y \in \left\{ 0,1,2, \dots, m \right\}$. As another example, the research may be interested in the response time of a person to a stimulus in an experimental task, in which case $y \in \mathbb{R}^+$. The possible set of values $y$ can assume is denoted as $S$ (also called the sample space).

In this paper, a statistical model is considered as a (possibly uncountable) collection of probability mass or density functions $\mathcal{M} = \left\{ p_1(y), p_2(y), \dots \right\}$ deemed suitable to describe the distribution of $y$. In what follows, I may use the shorthand term “distributions” referring to either a “probability mass functions” or a “probability density function”, depending on the context. 

\sloppy Any distribution makes predictions about a particular phenomenon in the world (as captured by $y$). Consequently, this means that any statistical model is in fact a mathematical or formalized theory about a particular aspect of the world. Such a theory may be very strong, in which case the set of models is a singleton: $\mathcal{M}_1 = \left\{ p(y) \right\}$. In this case, the only uncertainty is the irreducible uncertainty associated with not being able to tell which value for $y$ will be observed. Such a situation would mean that a researcher can be confident beyond any doubt about the  model for a particular phenomenon. This is a rather unlikely situation for behavioral scientists. Another extreme situation is that our set consists of all thinkable distributions for $y$: $\mathcal{M_{\infty}}=\left\{ p(y) \ | \ p(y) \geq 0 \mbox{ for all } y \in S \mbox{ and } \int_S p(y) dy = 1 \right\}$ (if $y$ is discrete, the integral should be interpreted as a sum). This is not really a desirable situation for at least two reasons. First, it may signal complete absence of knowledge about a phenomenon (aside from being able to define the random variable). Second, in case of continuous random variables, it is difficult to bring structure to the set $\mathcal{M}_{\infty}$ because it is an infinite dimensional space \cite{amari2016information}.

A very common situation (and also the focus of this paper) is that $\mathcal{M}$ consists of a so-called parametric family of distributions: $\mathcal{M}= \left\{ p(y|\theta) \ | \ \theta \in \Omega \right\}$. In this case, we are dealing with a statistical model of which the individual members are indexed by a parameter (or parameter vector) $\theta$. The parameter\footnote{The reader may wonder how the word "parameter" became part of the vocabulary of statistics. Despite the Greek origins of the word "parameter", its history is much more recent. The concept (not the word) of a parameter arises first in the work of Jordanus Nemorarius (1225-1260), who started working with letters (as opposed to specific numbers) thereby being able to treat larger sets of cases to which the same principles can be applied \cite{boyer2011history}. The word "parameter" has probably been introduced by French mathematician Claude Mydorge (1585-1648), who used it to refer to the latus rectum of a parabola \cite{harris1708lexicon,sugimoto2013discourse}. In statistics, it was Fisher who first used the term parameter to index a family of distributions \cite{stigler2005fisher}.} $\theta$ can be a scalar or a vector of dimension $k$. The set $\Omega \subseteq \mathbb{R}^k$ is called the parameter space. 

\begin{figure}
    \centering
    \includegraphics[width=0.9\linewidth]{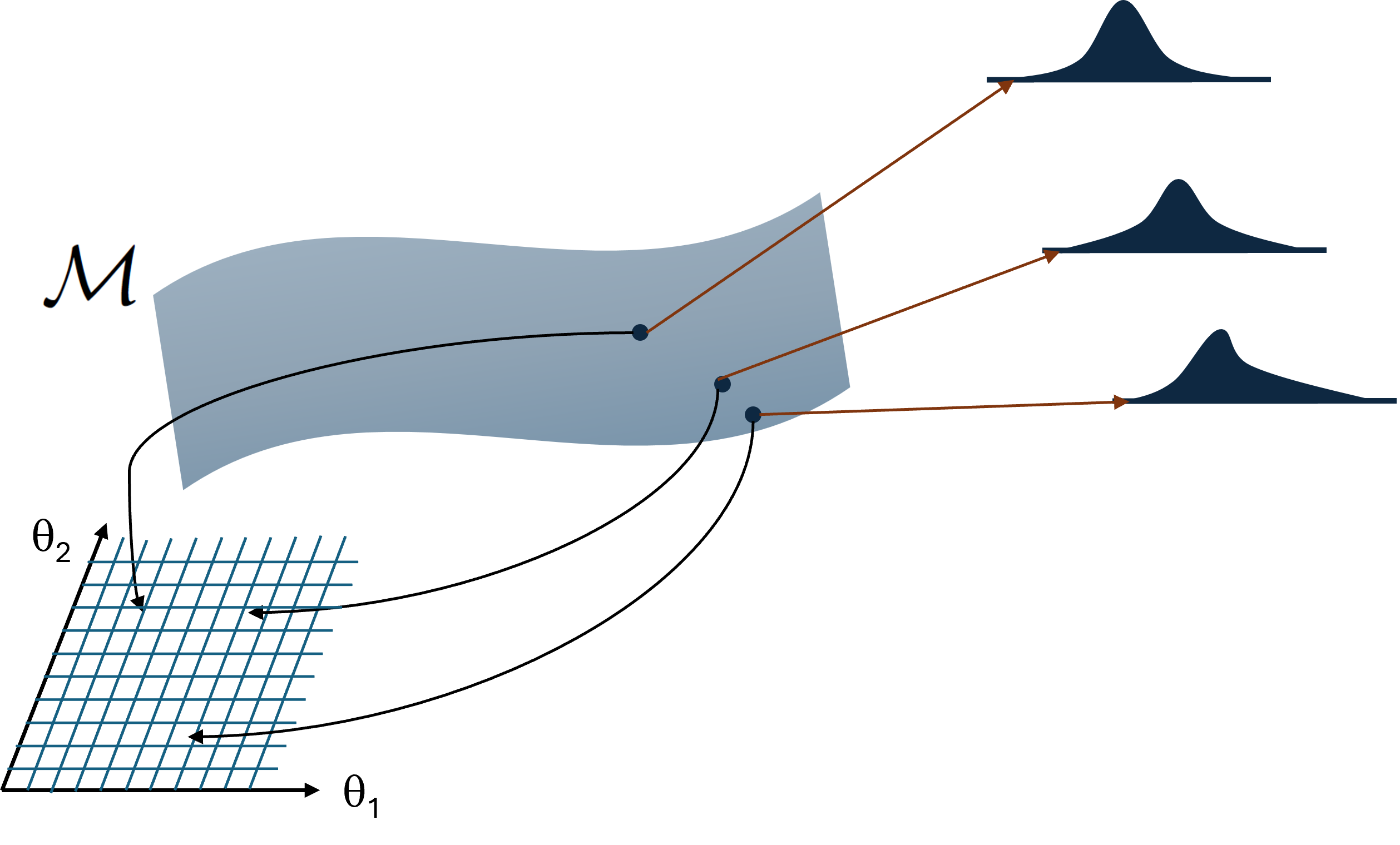}
    \caption{Graphical illustration of a smooth two-dimensional statistical manifold $\mathcal{M}$ together with a coordinate space $\Omega$. Each location on the manifold corresponds to a distribution of some family and the  $(\theta_1,\theta_2)$ is a pair of coordinates indexing the distributions.}
    \label{fig:manifolddistr}
\end{figure}

In what follows, we restrict our attention to the cases of regular parametric statistical models that constitute a smooth manifold of distributions (e.g., the manifold of univariate normals with differing location and scale). The two key concepts of the latter sentence will be explained next. First, a regular statistical model means that we require that the Fisher information (see below) is everywhere of full rank and that that model is well-identified model (i.e., $\theta \neq \theta^{\prime} \Rightarrow p(y|\theta) \neq p(y|\theta^{\prime})$). Second, a parametric statistical model $\mathcal{M}$ is a smooth manifold if an infinitely differentiable and invertible map exists from the manifold to the parameter space $\Omega \subseteq \mathbb{R}$ (this map is called a chart map or chart). This is illustrated graphically in Figure~\ref{fig:manifolddistr}. Note that in order to properly define such a smooth manifold, a number of priors steps have to be taken (such as defining a topological manifold) as well as a number of more technical issues need to be clarified (e.g., one has to deal with the situation that one chart map may not be sufficient to index all points on the manifold\footnote{For example, consider a stereographic projection from the sphere (i.e., the manifold) to a 2D plane. No projection succeeds in mapping all locations on the sphere onto a single 2D map. Therefore, a set of overlapping charts that cover the manifold (constituting an atlas) is used. Note that in this paper only global charts are used for statistical models, which means that one parametric coordinate system is sufficient to index all distributions.}). A key property of a smooth manifold is that it locally resembles a Euclidean space in the small neighborhood around every point (analogue to a first order Taylor approximation to a nonlinear function) and this will allow us to do vector calculus on the manifold. A smooth manifold of regular statistical distributions is called here a statistical manifold.

For the purpose of this paper, the presented superficial account of smooth manifolds suffices but interested readers can consult various sources for more information on smooth manifolds and differential geometry \cite<see e.g.,>{amari2016information,Boothby1986,tu2011manifolds,lee2003introduction}. The relation between statistics and differential geometry is more deeply elaborated in \citeA{amari2016information}, \citeA{calin2014geometric}, \citeA{kass1989geometry}, and \citeA{kass2011geometrical}. Two noteworthy papers from quantitative psychology with nice introductions into some of the material used in this paper are \citeA{ly2017tutorial} and \citeA{segert2019general}.

\begin{figure}
    \centering
    \includegraphics[width=0.9\linewidth]{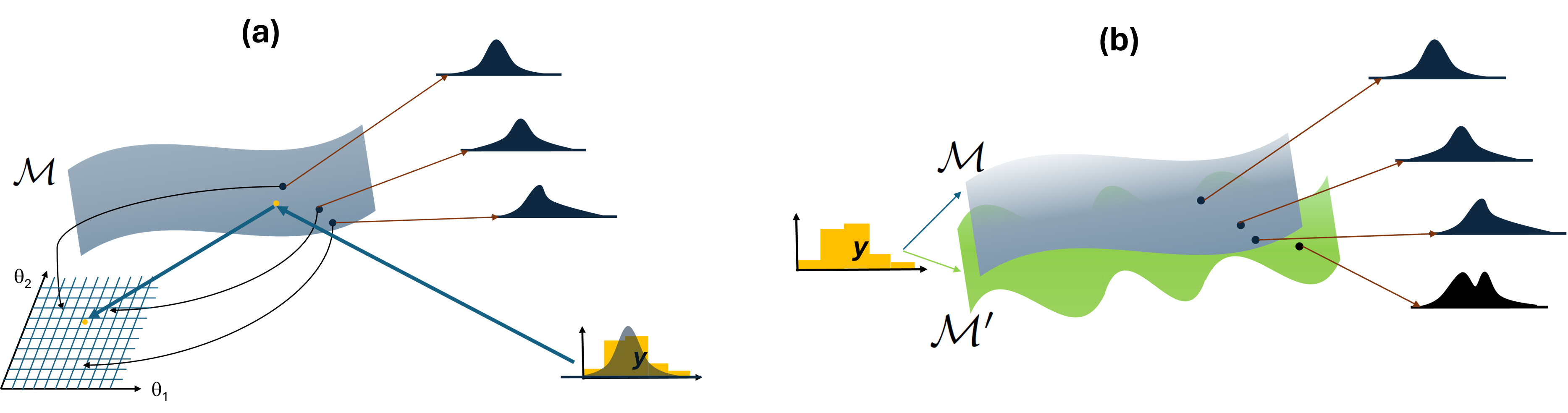}
    \caption{Graphical illustration of estimation (Panel (a)) and model selection (Panel (b)).}
    \label{fig:manifold_est_modsel}
\end{figure}

Formulating statistical models is one thing, bringing them into contact with empirical data and through this process of statistical inference learning about the world is another. Given observed data, an important task of (classical or frequentist) inferential statistics is selecting from $\mathcal{M}$, the distribution that is "closest" to the data in some sense. This is the domain of estimation (see Panel (a) in Figure~\ref{fig:manifold_est_modsel}). Because the members of $\mathcal{M}$ are indexed by the parameter (vector) $\theta$, estimation means parameter estimation, that is finding the optimal value of $\theta$ (often denoted as $\hat{\theta}$). An inalienable part of parameter estimation is assessing the uncertainty of the estimator $\hat{\theta}$. In another scenario, a second statistical model (also a parametric family) $\mathcal{M}'$ may be under consideration and then the question becomes which statistical model ($\mathcal{M}$ or $\mathcal{M}'$) fits the data best. This is the domain of model selection (see Panel (b) in Figure~\ref{fig:manifold_est_modsel}). The task of model selection is often easier when $\mathcal{M}'$ is nested within $\mathcal{M}$, in which case $\mathcal{M}' \subset \mathcal{M}$.

Alternatively, in the case of Bayesian statistics, one first defines a prior distribution $p(\theta|\mathcal{M})$ over the parameter space $\Omega$ and then derives (using Bayes' theorem) the posterior distribution $p(\theta |y,\mathcal{M})$. The (posterior) uncertainty about $\theta$ is intrinsically captured by its distribution. Likewise, we may assign prior probabilities to $\mathcal{M}$ and $\mathcal{M}'$ and derive their posterior probabilities $p(\mathcal{M}|y)$ and $p(\mathcal{M}'|y)$, thereby performing model selection.


The parameters of a statistical model play a crucial role in most studies. Usually, the interpretation of the results is based on one or more of the estimated parameters of interest. This is justified because these parameters are used to answer the question of interest (e.g., a difference between two conditions) or because the parameters represent a psychological process (e.g., speed of information accumulation) or attribute (e.g., a person's math ability). However, relying on parameters for scientific inference may also bring along a problem. The first and fundamental role of the parameters is to index the distributions. Parameters are only tools to identify the distributions that make  predictions about the world. In fact, as an indexing tool, there is a quite some degree of arbitrariness in the choice of parametrization. 

\paragraph{Example: The Rasch model}
To illustrate this point of arbitrariness of the choice of parametrization, we will make use of an example of \cite{ramsay1996geometrical}. Let us consider a Rasch model for a single person (with unknown ability) taking a test of $m$ items. We will assume that the items are fully known to us, so that we know the values of the $\beta_j$ item difficulties. The binary random variables $y_j$ (with 1 denoting success and 0 failure) can be collected in a vector $y$. The probability of success on item $j$ can be written now as follows:
\begin{equation} \label{eq:raschsingle}
\pi_j(\theta) = \frac{e^{\theta-\beta_j}}{1+e^{\theta-\beta_j}}  
\end{equation}
and the distribution of the response vector $y$ is the product-Bernoulli:
\begin{equation} \label{eq:raschjoint}
p(y|\theta) = \prod_{j=1}^m \pi_j(\theta)^{y_j} (1-\pi_j(\theta))^{1-y_j} = \prod_{j=1}^m\frac{e^{y_j(\theta-\beta_j)}}{1+e^{\theta-\beta_j}}.  
\end{equation}
The person specific ability parameter $\theta$ can be transformed without affecting the model predictions. For example, the following parametrizations for ability are equally valid: $\xi(\theta) = e^{\theta}$ ($\xi \in \mathbb{R}^+$) or $\psi(\theta) = 2\  \mbox{arctan}\left( e^{\frac{\theta}{2} } \right)$ (with $\psi \in (0,\pi)$ and $\mbox{arctan}$ being the arctangent or inverse tangent function). Both transformations are smooth invertible parameter transformations that do not affect the probability of a correct answer:
$$\pi_j(\theta)  
 =\pi_j(\xi)  =  \pi_j(\psi) 
$$
because
$$\frac{e^{\theta-\beta_j}}{1+e^{\theta-\beta_j}} 
   =  \frac{\xi e^{-\beta_j}}{1+\xi e^{-\beta_j}} 
  = \frac{\tan^2 \left( \frac{\psi}{2} \right) e^{-\beta_j}}{1+\tan^2 \left( \frac{\psi}{2}\right) e^{-\beta_j} }.
$$
This problem of the parameter-dependent and thus arbitrary ability scale in item response theory has been discussed by \citeA{ramsay1996geometrical}. As will be shown below, the solution proposed by \citeA{ramsay1996geometrical} to arrive at a parametrization-invariant scale also rests on geometrical arguments (although he uses classical differential geometry, which leads to results somewhat different from mine, see below). \demo\footnote{This symbol is used in this paper to mark the end of an example.}
In the following sections, I will discuss three important concepts from geometry and their relation to psychometric and quantitative psychology: Distance, curvature and volume.

\section{Distance}

A first geometrical property I will discuss is distance. Because we are dealing with distributions on a manifold, the goal is to study the distance between two distributions of the smooth manifold $\mathcal{M}$. To obtain the reasoning behind a meaningful distance measure, I will follow the original line of reasoning of \citeA{rao1945} (see also \citeNP{catichaEntropic}). Start with two nearby points $\theta$ and $\theta+d\theta$ in the parameter space $\Omega$. In the Euclidean parameter space $\Omega$, we can make use of Pythagoras' theorem to find the distance between these two points: $d(\theta,\theta+d\theta)^2 = \sum_i d\theta_i^2 = d\theta^T d\theta$ (where the subscript $T$ is the transpose).

A reasonable question to ask is whether we can take $d(\theta,\theta+d\theta)$ as the distance between the two distributions. To find the answer to this question, we will start out with the relative difference or deviation $\Delta$ between the two corresponding distributions (it will become clear below why this relative difference is a natural starting point):
\begin{equation} \label{eq:deltay}
\Delta(y) = \frac{p(y|\theta+d\theta)-p(y|\theta)}{p(y|\theta)},    
\end{equation}
where $\Delta$ depends on $y$. Using a first-order Taylor series approximation to, $p(y|\theta+d\theta)\approx p(y|\theta) + \sum_i \frac{\partial p(y|\theta)}{\partial \theta_i} d\theta_i$, and plugging the result into Equation~\ref{eq:deltay}, gives:
$$\Delta(y) \approx \frac{1}{p(y|\theta)} \sum_i \frac{\partial p(y|\theta)}{\partial \theta_i} d\theta_i = \sum_i \frac{\partial \log p(y|\theta)}{\partial \theta_i} d\theta_i.$$
To eliminate the dependency on $y$, the expectation of $\Delta(y)$ with respect to $p(y|\theta)$ is taken. However, $E\left[ \Delta(y) \right]=0$ and thus $\Delta(y)$ is not a good basis for a distance measure. 

However, using the squared relative difference $\Delta(y)^2$ proves to be a more viable path. If we then take the expected value we have the mean squared relative deviation of the distributions, which is also called distinguishability:
\begin{align}
E\left[ \Delta(y)^2 \right] &= \int_S \left( \frac{1}{p(y|\theta)} \sum_i \frac{\partial p(y|\theta)}{\partial \theta_i} d\theta_i \right)^2 p(y|\theta) dy \nonumber \\
&= \int_S \left( \frac{1}{p(y|\theta)} \sum_i \frac{\partial p(y|\theta)}{\partial \theta_i} d\theta_i \right) \left( \frac{1}{p(y|\theta)} \sum_j \frac{\partial p(y|\theta)}{\partial \theta_j} d\theta_j \right) p(y|\theta) dy \nonumber \\
&=  \int_S \left( \sum_i \frac{\partial \log p(y|\theta)}{\partial \theta_i} d\theta_i \right) \left(  \sum_j \frac{\partial \log p(y|\theta)}{\partial \theta_j} d\theta_j \right) p(y|\theta) dy \nonumber \\
&= \sum_{i,j} E \left[ \left( \frac{\partial \log p(y|\theta)}{\partial \theta_i} \right) \left(  \frac{ \partial \log p(y|\theta)}{\partial \theta_j} \right) \right] d\theta_i d\theta_j \nonumber \\
&= ds^2, \label{eq:fisherexplog}
\end{align}
where we have used the notation $ds^2$ to indicated that this mean squared relative deviation is considered as the squared distance between two distributions ont he manifold. The expected value of the product of first derivatives of the log density in Equation~\ref{eq:fisherexplog} is the $(i,j)$th element of the Fisher information matrix $g(\theta)$. In matrix notation:
$$g(\theta)= E \left[ \frac{\partial \log p(y|\theta)}{\partial \theta^T}\frac{\partial \log p(y|\theta)}{\partial \theta} \right]. \label{eq:fishermatrix} $$
For the statistical models considered in this paper, it can be shown $g_{ij}(\theta) =-E \left[ \frac{\partial^2 \log p(y|\theta)}{\partial \theta_i \partial \theta_j} \right]$ (see \citeNP{calin2014geometric}).

As a result, we have for the squared distance element $ds^2$ :
\begin{equation}
ds^2  =  \sum_i \sum_j g_{ij}(\theta) d\theta_i d\theta_j = d\theta^T g(\theta) d\theta. \label{eq:frmetric}
\end{equation}
This can be considered a generalization of the well-known Pythagorean theorem for Euclidean geometry. In general, $ds^2 \neq d(\theta,\theta+d\theta)^2$, unless $g(\theta)=1$ for all $\theta$, in which case we are back in the Euclidean case. This derivation shows why the (squared) relative distance was a natural starting point. 

The squared distance element $ds^2$ informs us about the local distinguishability between $p(y|\theta)$ and its nearby distributions. A more common quantity to assess differences between distribution is the Kullback-Leibler divergence $D_{KL}$. Using $D_{KL}$, the difference of $p(y|\theta')$ from $p(y|\theta)$ is:
$$
D_{KL}\left( p(y|\theta) \Vert p(y|\theta') \right) = \int_S p(y|\theta) \log \left( \frac{p(y|\theta)}{p(y|\theta')} \right) dy.
$$
Although $D_{KL} \geq 0$ (with equality when $p(y|theta)=p(y|theta')$ for all $y$), it is not symmetric, nor satisfies the triangle inequality, and thus the $D_{KL}$ divergence is not a distance. However, when $\theta'=\theta+d\theta$, the quadratic approximation to $D_{KL}$ becomes:
$$
D_{KL}\left( p(y|\theta) \Vert p(y|\theta+d\theta) \right) \approx \frac{1}{2} ds^2,
$$
which means that for nearby distributions, the $D_{KL}$ divergence is actually carrying the same information as the local distinguishibility $ds^2$.

The Fisher information matrix is also called the Fisher-Rao metric. It was Rao's (1945) insight that Fisher's information matrix $g(\theta)$ is in fact a so-called metric tensor\footnote{More exactly, it is a (0,2) metric tensor, also called 2-covariant tensor.} that is used to measure distances on smooth manifold equipped with this metric. A differentiable manifold endowed with such a metric is called a Riemannian manifold. 

\begin{figure}
    \centering
\picdims[width=0.5\linewidth]{6cm}{6cm}{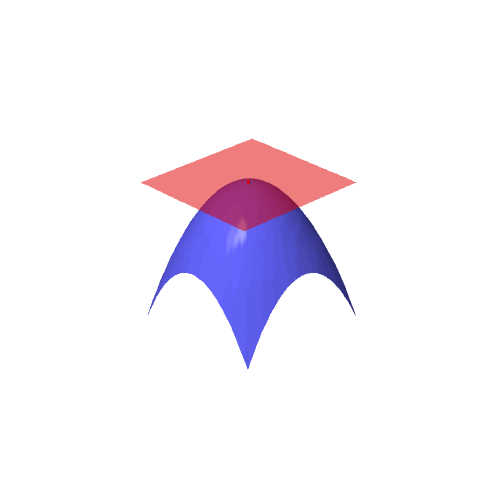}
    \caption{Illustration of the tangent space $T_{p_0}\mathcal{M}$ (red plane) to the manifold $\mathcal{M}$ (blue surface) at the point $p_0 \equiv p(y|\theta_0)$ (red dot).}
   \label{fig:tangent}
\end{figure}

In fact, at each point $p(y|\theta_0) \equiv p_0$  of the $k$-dimensional manifold $\mathcal{M}$, a a tangent subspace $T_{p_0}\mathcal{M}$ can be defined. If the manifold is embedded in the Euclidean space, then the tangent space is the best linear approximation to the manifold in the point of approximation (see Figure~\ref{fig:tangent} for an illustration). This tangent space $T_{p_0}\mathcal{M}$ is defined by all tangent vectors attached to the point $p(y|\theta_0)$. $T_{p_0}\mathcal{M}$ is a vector space endowed with an inner product, which can be used to calculate norms of tangent vectors and distances (see below).


Using Equation~\ref{eq:frmetric}, we have a generalization of Pythagoras' theorem for non-Euclidean curved spaces such as our smooth manifold $\mathcal{M}$. The squared distance $ds^2$ (and thus also the distance $ds$) does not change under parameter transformations. Assume that $\phi = \phi(\theta)$ and $\theta=\theta(\phi)$ such that $\phi(\cdot)$ is one-to-one and $\phi(\cdot)$ and $\theta(\cdot)$ are smooth (i.e., the transformation is a diffeomorphism). Then we can write that:
$$
d s^2 =\sum_i \sum_j g_{ij}(\theta) d\theta_i d\theta_j = \sum_i \sum_j g^*_{ij}(\phi) d\phi_i d\phi_j,
$$
where $g^*_{ij}(\phi)$ is the Fisher information matrix under the new parametrization $\phi = \phi(\theta)$. For a general proof for the parameter vector case, we refer to \citeA{calin2014geometric}). For the univariate case, it can be demonstrated as follows. First, define the following: $p^*(y|\phi)=p(y|\theta(\phi))$. Second, let us use the chain rule: $\frac{d\log p^*(y|\phi)}{d\phi} = \frac{d\log p(y|\theta(\phi))}{d\phi}  = \frac{d\log p(y|\theta)}{d\theta} \frac{d\theta}{d\phi}$. Using this in the definition of the Fisher information gives:
\begin{align} 
g^*(\phi) &= \int_S p^*(y|\phi) \left( \frac{d\log p^*(y|\phi)}{d\phi}\right)^2 dy \nonumber \\ 
&= \int_S p(y|\theta(\phi))  \left( \frac{d\log p(y|\theta)}{d\theta}\frac{d\theta}{d\phi}\right)^2 dy \nonumber \\
&= g(\theta) \left(\frac{d\theta}{d\phi}\right)^2. \label{eq:transformfisher}
\end{align}
Because the differential $d\theta$ transforms as well, $d\theta=\frac{d\theta}{d\phi}d\phi$, the squared distance element remains constant:
$$ 
ds^{*2} = g^*(\phi) (d\phi)^2 = g(\theta) \left( \frac{d\theta}{d\phi} \right)^2 (d\phi)^2 = g(\theta) (d\theta)^2 = ds^2.
$$

Although $ds$ is a distance element, it is a dimensionless quantity. To understand this, assume for simplicity that $\theta$ is a scalar parameter. As is well-known from statistics, the inverse of Fisher information $g(\theta)$ is the variance of an estimator for the parameter $\theta$. Therefore, the dimension of $g(\theta)$ is equal to the dimension of $\theta^{-2}$. Taken together with $d\theta^2$ (having dimension of $\theta^{2}$), this results in a dimensionless $ds^2$ and thus also a dimensionless $ds$.

A distance on the manifold $\mathcal{M}$ between two arbitrary distributions $p(y|\theta)$ and $p(y|\theta')$ can then be calculated by defining a path between them and then computing the length of the path (a so-called arc length). Assume a parametrized curve $\theta(t)$ with $t \in [t_0,t_1]$ such that  $\theta(t_0)=\theta$ and $\theta(t_1)=\theta'$. The arc length on the manifold $\mathcal{M}$ is found by assuming a fine-grained grid along the path and then accumulating the many small distance elements $ds$ (and also taking the absolute value to ensure a positive distance):
\begin{equation} \label{eq:arclength}
d_{\theta(t)}( p(y|\theta), p(y|\theta') ) = \int_{t_0}^{t_1} \sqrt{ \sum_i \sum_j g_{ij} \frac{d\theta_i(t)}{dt} \frac{d\theta_j(t)}{dt}} dt.  
\end{equation}
Obviously, another path $\theta^*(t)$ will likely lead to another arc length. One particular important path is the one with the shortest possible arc length. This is the so-called geodesic curve, denoted here as $\theta_{\min}(t)$:
$$
\theta_{\min}(t) = \arg \min_{\theta(t)} d_{\theta(t)}( p(y|\theta), p(y|\theta') ).
$$
Plugging in this curve in Equation~\ref{eq:arclength} gives the geodesic distance $d_{\min}$, which is the shortest arc length between the two distributions $p(y|\theta)$ and $p(y|\theta')$:
$$
d_{\min}( p(y|\theta), p(y|\theta') ) = \min_{\theta(t)} d_{\theta(t)}( p(y|\theta), p(y|\theta') ).
$$
The geodesic (or Fisher-Rao) distance is an invariant and does not depend on the parametrization. Moreover, it is a genuine distance (using $p$, $q$ and $r$ for distributions): (1) $d_{\min}(p,q)\geq 0$ with $d_{\min}(p,p)=0$ (non-negativity), (2) $d_{\min}(p,q) = d_{\min}(q,p)$ (symmetry), and (3) $d_{\min}(p,q)+d_{\min}(q,r) \geq d_{\min}(p,r)$ (triangle inequality) \cite{lee2003introduction}.

Finding the geodesic curve is in most situations not straightforward and requires the solution to a variational problem \cite<see e.g.,>{atkinson1981rao,calin2014geometric}. For a number of common distributions, these geodesic (or Fisher-Rao) distances have been derived and can be found in various sources, \cite<see e.g.>{miyamoto2024closedform,atkinson1981rao,calin2014geometric}. I will discuss two examples below: the normal distribution and the Rasch model.

\paragraph{Example: Normal distribution}

To illustrate some of the aforementioned concepts, let us consider the very common normal distribution $N(\mu,\sigma^2)$ with density $p(y|\theta) = \frac{1}{\sqrt{2\pi \sigma^2}}e^{-\frac{1}{2}\frac{(y-\mu)^2}{\sigma^2}}$ with parameters $\theta=(\mu,\sigma)$. The parameter space $\Omega$ is the upper half plane: $\Omega = \mathbb{R} \times \mathbb{R}^+$. The Fisher information matrix is:
$$ \label{eq:gnormal}
g = 
\begin{pmatrix}
\frac{1}{\sigma^2} & 0 \\
0 & \frac{2}{\sigma^2} 
\end{pmatrix}.
$$
Using Equation~\ref{eq:frmetric}, the squared distance element for the normal distribution can be written as:
$$
ds^2 = \frac{(d\mu)^2+2(d\sigma)^2}{\sigma^2}.
$$

\begin{figure}
    \centering
   \includegraphics[width=0.6\linewidth]{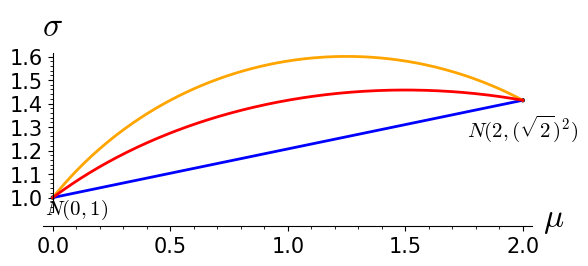}
    \caption{Three possible curves between $N(0,1)$ and $N\left(2,(\sqrt{2})^2 \right)$: A straight line (in blue), a circular arc (in orange) and an ellipse (in red, also the geodesic curve). The distances along the three curves are 1.744, 1.697, and 1.656, respectively. See text for the parametric equations and the calculations.}
    \label{fig:curvesNormal}
\end{figure}

In a next step, let us calculate the distance between the standard normal $N(0,1)$ and $N(2,(\sqrt{2})^2)$ along three paths (see Figure~\ref{fig:curvesNormal}). We do this by first considering a straight line in the parameter space from $\theta_0=(\mu_0,\sigma_0)= (0,1)$ to $\theta_1=(\mu_1,\sigma_1)= (2,\sqrt{2})$. Such a straight line can be parametrized by the following parametric equation: $\theta_{\vert}(t) = (\mu(t),\sigma(t)) = (2t,1+t(\sqrt{2}-1))$ with $t \in [0,1]$. It is easy to check that $\theta_{\vert}(0)=(0,1)=(\mu_0,\sigma_0)$ and $\theta_{\vert}(1)=(2,\sqrt{2})=(\mu_1,\sigma_1)$. Moreover, $\frac{d\theta_{\vert}(t)}{dt}=\left( \frac{d\theta_{1,\vert}(t)}{dt},\frac{d\theta_{2,\vert}(t)}{dt} \right)=\left( \frac{d\mu(t)}{dt},\frac{d\sigma(t)}{dt} \right)=(2,\sqrt{2}-1)$. Inserting this into Equation~\ref{eq:arclength} gives\footnote{From now on, the absolute value vertical bars have been dropped because I make sure the paths are defined in such a way that the result is positive.}:
\begin{align*}
    d_{\vert} &= \int_0^1 \sqrt{\frac{2^2+2(\sqrt{2}-1)^2}{1+t(\sqrt{2}-1)^2} } dt  = \sqrt{10-4\sqrt{2}} \int_0^1 \frac{1}{1+t(\sqrt{2}-1)} dt \\
    &=\frac{\log(2)}{2} \frac{ \sqrt{10-4\sqrt{2}}}{\sqrt{2}-1} \approx 1.744.
\end{align*}

Without comparison it is hard to interpret this number. For that reason, let us define a circular arc between $(0,1)$ and $(2,\sqrt{2}))$ with the center of the circle lying on the $\mu$-axis. This circle passing through these points is centered at $(\frac{5}{4},0)$ with radius $\frac{\sqrt{41}}{4}$ and has the following parametric equation: $\theta_{\circ}(t) = (\mu(t),\sigma(t)) = \left( \frac{5}{4}+\frac{\sqrt{41}}{4}\cos t,\frac{\sqrt{41}}{4}\sin t \right)$. If we restrict $t$ to lie in the interval $[t_0,t_1] = \left[ \cos^{-1}\left(  \frac{3}{\sqrt{41}} \right), \cos^{-1}\left(  -\frac{5}{\sqrt{41}} \right) \right]$ then we obtain the desired circular arc (see Figure~\ref{fig:curvesNormal}). Subsequently, $\frac{d\theta_{\circ}(t)}{dt}=\left( \frac{d\mu(t)}{dt},\frac{d\sigma(t)}{dt} \right)=\left( -\frac{\sqrt{41}}{4} \sin t,\frac{\sqrt{41}}{4} \cos t \right)$. Inserting this information into Equation~\ref{eq:arclength} (with appropriate integration limits) results in:
\begin{align*}
    d_{\circ} &=  \int_{t_0}^{t_1} \sqrt{\frac{\frac{41}{16} \sin^2 t + 2\frac{41}{16} \cos^2 t}{\frac{41}{16} \sin^2 t} } dt  = \int_{t_0}^{t_1} \frac{\sqrt{\sin^2 t + 2\cos^2 t}}{\sin t} dt\\
    &=\log \frac{2\sqrt{33}-5\sqrt{2}}{10+3\sqrt{2}} +\sqrt{2}\left(\tanh^{-1}\left(\frac{3}{5}\right)+\tanh^{-1}\left(\frac{5}{\sqrt{33}}\right) \right) \approx 1.697.
\end{align*}
Thus paradoxically, the circular arc leads to shorter distance between the two normals than the straight line.

However, to find the shortest distance between the two normals $N(0,1)$ and $N(2,(\sqrt{2})^2)$, the geodesic distance can be computed. As a first step, the geodesic path needs to be computed as is explained in \citeA{calin2014geometric,GelmanMeng1998,atkinson1981rao}. Based on the geodesic path, the distance can be computed based on Equation~\ref{eq:arclength}. In \citeA{costa2015fisher}, closed-form formulas for the normal distribution are presented. Let us start with this formula:
\begin{multline*}
d_{\min}( (\mu_0,\sigma_0),(\mu_1,\sigma_1) )  = \sqrt{2} \log \frac{ \splitdfrac{\sqrt{\left( (\mu_0-\mu_1)^2 +2(\sigma_0-\sigma_1)^2\right) \left((\mu_0-\mu_1)^2 +2(\sigma_0+\sigma_1)^2 \right) }}{+(\mu_0-\mu_1)^2+2(\sigma^2_0+\sigma^2_1)}}{4\sigma_0 \sigma_1}.
\end{multline*}
For our example, $d_{\min} = \sqrt{2} \log \frac{\sqrt{17}+5}{2\sqrt{2}}\approx 1.656$. As expected, $d_{\min} < d_{\circ} < d_|$. This then raises the questions how the geodesic path looks like. It is the arc of an ellipse centered at $(1.5,0)$ and parametric equation\footnote{What follows is the equation of an ellipse with width $\sqrt{17}$ and height $\sqrt{\frac{17}{2}}$.}: $\theta(t) = \left( \frac{3}{2} + \frac{\sqrt{17}}{2} \cos t, \frac{\sqrt{17}}{2\sqrt{2}} \sin t\right)$. For the elliptic arc, $t \in \left[ \sin^{-1}\left(\frac{4}{\sqrt{17}}\right), \cos^{-1}\left(-\frac{3}{\sqrt{17}}\right)\right]$. Inserting these ingredients into Equation~\ref{eq:arclength}, gives (once more): $d_{\min} \approx 1.656$. 

\begin{figure}
    \centering
    \includegraphics[width=0.4\linewidth]{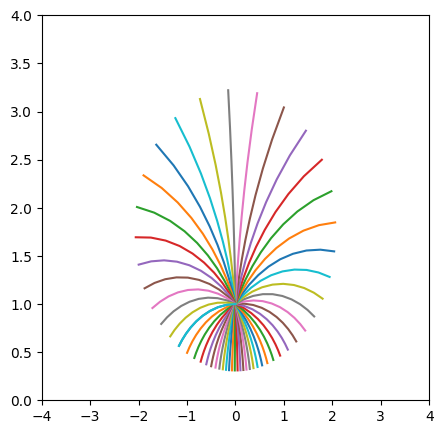} \hfill
    \includegraphics[width=0.4\linewidth]{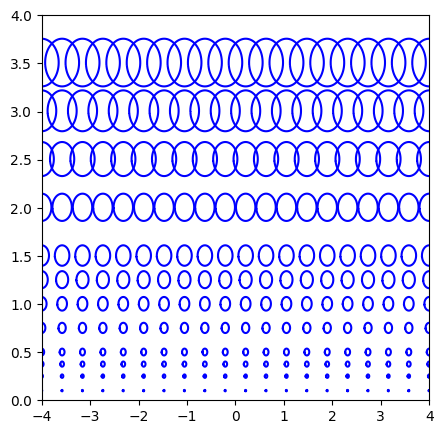}
    \caption{The left panel shows 50 geodesic curves (as rays) all starting at $N(0,1)$ and ending at the distribution that is $d_{\min}=1.656$ away. The right panel shows equidistant points at 0.01 from the center for various combinations of $\mu$ and $\sigma$. These figures are created using the Python package "geomstats" (Miolane et al., 2020). }
    \label{fig:geodesicballnormal}
\end{figure}

In the left panel of Figure~\ref{fig:geodesicballnormal} a number of geodesic paths, all emanating from $N(0,1)$, are shown. Every path has the same the starting point $N(0,1)$ and every path ends at a distance $d_{\min}=1.656$ from $N(0,1)$. The different geodesic rays show the set of normal distributions that are equidistant from the standard normal. Such a set of equidistant points based on geodesics is called a geodesic ball. The right panel of Figure~\ref{fig:geodesicballnormal} shows for several $(\mu,\sigma)$ pairs the geodesic balls of radius 0.01. \demo

\paragraph{Example: The Rasch model}
The results for the normal distribution are well-known and described already many times. Let us now consider a lesser known example: a Rasch model for a single person (with unknown ability) taking a test of $m$ items. As before, we will assume that the item difficulties $\beta_j$ are known. The probability correct response on a single item $j$ is given by Equation~\ref{eq:raschsingle} and the joint probability to the $m$ items by Equation~\ref{eq:raschjoint}. This is a one-dimensional manifold $\mathcal{M}$. As is well-known, the Rasch model is a one-parameter exponential family model \cite{Efron_2023}:
\begin{align*}
p(y|\theta) &= \exp \left( y_+ \theta  - \sum_j \log \left( 1+e^{\theta-\beta_j}\right) \right)\exp \left( - \sum_j y_j \beta_j \right) 
\end{align*}
with $\theta$ as the natural parameter, $y_+$ the sufficient statistic, $\Psi(\theta)=\sum_j y_j \beta_j - \sum_j \log \left( 1+e^{\theta-\beta_j}\right)$ (the log-partition function), and $g_0(y) = \exp \left( - \sum_j y_j \beta_j \right)$ the so-called carrying density. Because it is a one-parameter model, the Fisher information matrix reduces to a scalar:
\begin{align*}
g &= E\left( -\frac{\partial^2 \log p(y|\theta)}{\partial \theta^2} \right) = \frac{\partial^2 \Psi(\theta)}{\partial \theta^2} \\
&= \sum_j \pi_j(\theta)(1-\pi_j(\theta)) = \sum_j \frac{e^{\theta-\beta_j}}{\left(1+ e^{\theta-\beta_j}\right)^2} 
\end{align*}
The Fisher information is in this context also called the test information function (because it gives the amount of information provided by the test of $m$ items). The squared distance element equals:
$$
ds^2 = g(\theta) (d\theta)^2.
$$

A natural question to ask is what in this case the geodesic distance is between two test takers with parameter values $\theta_0$ and $\theta_1$ (with $\theta_0 <\theta_1$). To find this, we follow the argument of \cite{atkinson1981rao}. Because $\mathcal{M}$ is unidimensional, a single parameter is sufficient to index all distributions. Our goal is to find a transformation that maps $\theta$ into $\phi$, such that the metric of $\phi$ is that of a Euclidean space (such a map exists for unidimensional manifolds, but in general not for higher dimensional ones):
$$
ds^2 = g^*(\phi) (d\phi)^2 = (d \phi)^2,
$$
because $g^*(\phi) =1$ for all $\phi$. In that case, the geodesic distance between two test takers can be simply computed as $\phi_1-\phi_0$. Expressing this difference as a function of the familiar $\theta$ then gives the solution.  

How to find the map from $\theta$ to $\phi$ (i.e., $\phi = \phi(\theta)$) so that the result is a Euclidean metric? If for a parametrization $\phi$, the metric is Euclidean, this means that: $ds^2 = (d\phi)^2$. This means that we require $g^*(\phi) = 1$. Using the transformation of the Fisher information then gives:
\begin{align*}
    1 &= g(\theta) \left( \frac{d\theta}{d\phi }\right)^2 \\
    1 &= \sqrt{g(\theta)} \frac{d\theta}{d\phi } \\
    d\phi &= \sqrt{g(\theta)}  d\theta \\
    \phi &= \int \sqrt{g(\theta')} d\theta' + C,
\end{align*}
where we have only considered the positive square root and have used the name $\theta'$ as the variable with which to integrate (in order to distinguish it from $\theta$). For simplicity, we will set $C=0$ because the constant will cancel when taking differences.

For the ability $\theta_i$ of person $i$ ($i=0,1$), we can now compute the corresponding $\phi_i = \phi(\theta_i) =  \int_L^{\theta_i} \sqrt{g(\theta')} d\theta'$. The geodesic distance $d_{\min}(\theta_0,\theta_1)$ can then be calculated:
$$
d_{\min}(\theta_0,\theta_1) = \phi_1-\phi_0 =  \int_{\theta_0}^{\theta_1} \sqrt{g(\theta')} d\theta' . 
$$
Inserting the expression for the Fisher information (or test information function) gives:
$$ \label{eq:rasch1Dgeod}
d_{\min}(\theta_0,\theta_1) = \int_{\theta_0}^{\theta_1} \sqrt{\sum_j \frac{e^{\theta'-\beta_j}}{\left(1+ e^{\theta'-\beta_j}\right)^2}} d\theta' .
$$
The geodesic distance $d_{\min}(\theta_0,\theta_1)$ is independent of the specific parametrization and will be the same under any diffeomorphic transform of $\theta$. The geodesic distance can be used to define a geodesic ability $A(\theta)$:
$$
A(\theta) = d_{\min}(-\infty,\theta) =  \int_{-\infty}^{\theta} \sqrt{\sum_j \frac{e^{\theta'-\beta_j}}{\left(1+ e^{\theta'-\beta_j}\right)^2}} d\theta' .
$$
The geodesic ability scale $A(\theta)$ is parameter-invariant. Consider a smooth transformation $\tau=\tau(\theta)$ (using Equation~\ref{eq:transformfisher}, but with $\tau$ instead of $\phi$):
$$
A(\tau) =  \int^{\tau}_{-\infty} \sqrt{g^*(\tau')} d\tau' =  \int^{\theta}_{-\infty} \sqrt{g(\theta')}\frac{d \theta'}{d \tau'} d\tau'  =  \int^{\theta}_{-\infty} \sqrt{g(\theta')}d \theta' = A(\theta).
$$
Moreover, the geodesic ability $A(\theta)$ has an absolute zero: $A(-\infty)=0$ and is unbounded from above.

The geodesic ability $A(\theta)$ corresponds quite well with the requirements set out by \cite{vandermaas2011cognitive} for ability: Abilities are essentially positive, but can be absent as well (which corresponds to 0 on the ability scale). It is also striking that \citeA{vandermaas2011cognitive} illustrate the concept of an ability with the ability to walk, which "refers to a capacity to do something, namely, to cover a certain distance by using a particular form of propulsion common to land animals" \cite[p.344]{vandermaas2011cognitive}. One could say that this exactly what $A(\theta)$ represents: A distance walked by the test taker through the statistical model manifold $\mathcal{M}$. 

The geodesic ability $A(\theta)$ is also a refinement of the idea by \citeA{ramsay1996geometrical}, who takes a (classical differential) geometrical view on the item response theory by considering the probability curve $(\pi_1(\theta),\pi_2(\theta),\dots, \pi_m(\theta))$ in the $m$-dimensional (Euclidean) space and measuring its arc length. For the Rasch model, the arc length proposed by \citeA{ramsay1996geometrical} is:
$$
s(\theta) =   \int_{-\infty}^{\theta} \sqrt{\sum_j \frac{e^{2(\theta'-\beta_j)}}{\left(1+ e^{\theta'-\beta_j}\right)^4}} d\theta' .
$$
The two distances $s(\theta)$ and $A(\theta)$ are not equal but share some similarities and a closer study should be considered. A major difference is that the arc length in \citeA{ramsay1996geometrical} is calculated in Euclidean space while ours is on the manifold, taking into account the curved space (see below).

For the general integral in Equation~\ref{eq:rasch1Dgeod} to calculate $A(\theta)$, there exists (to the best of my knowledge) no closed form solution. However, we can find some analytical results by considering a simple case: Suppose the test has $m$ items of difficulty all equal to 0. For that particular case, the geodesic ability, denoted now as $A_0(\theta)$, simplifies to (we can ignore the absolute values symbols):
\begin{align*}
A_0(\theta) &= \int_{-\infty}^{\theta} \sqrt{\sum_j \frac{e^{\theta'}}{\left(1+ e^{\theta'}\right)^2}} d\theta' \\
&= \sqrt{m}  \int_{-\infty}^{\theta} \sqrt{\frac{e^{\theta'}}{\left(1+ e^{\theta'}\right)^2}} d\theta' \\
&= 2\sqrt{m} \tan^{-1}\left( e^{\frac{\theta}{2}} \right).
\end{align*}
Note that this result is actually equal to the geodesic distance between binomial distributions as derived by \citeA{atkinson1981rao}. It can be seen that $\lim_{\theta \rightarrow -\infty} A_0(\theta) = 0$ and $\lim_{\theta \rightarrow \infty} A_0(\theta) = \pi \sqrt{m}$.

\begin{figure}
    \centering
    \includegraphics[width=1\linewidth]{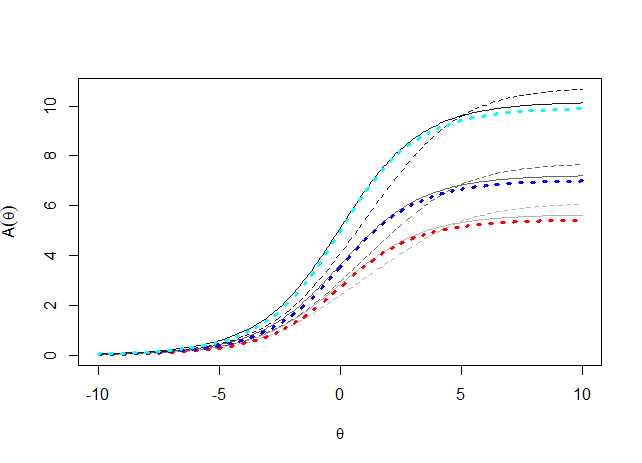}
    \caption{The geodesic ability $A(\theta)$ (or $A_0(\theta)$) as a function of $\theta$. The colored thick dotted lines are $A_0(\theta)$ for $m=3$ (red), $m=5$ (blue) and $m=10$ (cyan). The solid lines are have (from bottom to top) the following $\beta$ values (denoted as $a(b)c$, such that a sequence starts at $a$, ends at $b$ and has $c$ units between consecutive values): $(-1(1)1)$, $(-1(0.5)1)$, $(-1(2/9)1)$. The dashed lines have asymmetrically placed $\beta$ values: $(-1(2)3)$, $(-1(1)3)$, $(-1(4/9)3)$.}
    \label{fig:geodesicAbilityPlot}
\end{figure}

Figure~\ref{fig:geodesicAbilityPlot} graphically illustrates $A(\theta)$ and $A_0(\theta)$ as functions of $\theta$ for various choices of $\beta$ and number of items $m$. Several observations can be made from this plot. First, the plots show that the geodesic ability starts at 0 and is positive. Second, the major factor that determines where the function levels off is the number of items. Third, if the items are placed symmetrically around zero, $A_0(\theta)$ serves as an approximation to $A(\theta)$. 
 
Suppose one has to estimate the geodesic ability of a person based on a response vector. The standard IRT software can be used to obtain an estimator of $\theta$, denoted as $\hat{\theta}$. Plugging this estimate into the $A(\theta)$ function, then gives the estimated geodesic ability: $\widehat{A(\theta)} = A(\hat{\theta})$. It is also important to quantify the uncertainty of this estimator. To study this, let us assume again a test of $m$ items with all $\beta_j=0$. Thus, we want to assess the uncertainty of $A_0(\hat{\theta})$. From statistical theory, we know that $\sqrt{m}(\hat{\theta}-\theta) \xrightarrow{D} N\left( 0, m \cdot g^{-1}(\theta) \right)$, where $g^{-1}(\theta)$ is the reciprocal of the Fisher information (in the multiparameter case, it is the inverse of the Fisher information matrix). We multiply $g^{-1}$ by $m$ to offset the $\sqrt{m}$ factor at the left hand side (because $g$ is the Fisher information of $m$ random variables, not of a single one). To apply the delta method, we first computed $A_0'(\theta)=\frac{dA_0(\theta)}{d\theta} = 2\sqrt{m}\frac{e^{\frac{\theta}{2}}}{1+e^{\theta}}$. This then gives:
\begin{align*}
\sqrt{m}\left( A_0(\hat{\theta}) - A_0(\theta) \right) &\xrightarrow{D} N\left( 0, m g^{-1}(\theta) \left( A_0'(\theta) \right)^2 \right) \\
&\xrightarrow{D} N\left( 0, m \frac{(1+e^{\theta})^2}{m e^{\theta}}\left( 2\sqrt{m}\frac{e^{\frac{\theta}{2}}}{1+e^{\theta}} \right)^2 \right) \\
&\xrightarrow{D} N(0,4m).
\end{align*}
Consequently, $SE\left( A_0(\hat{\theta}) \right) \approx 2$. Hence, we have an estimator with a constant uncertainty. This will not hold exactly for the general $A(\hat{\theta})$, but it can be expected to hold approximately.

It is interesting to note that this result coincides with the findings by \citeA{hougaard1982parametrizations}. Based on earlier work, \citeA{hougaard1982parametrizations} considers the following transformation in an exponential family model:
$$
B_{\delta}(\theta) = \int_L^{\theta} \left[ \frac{d^2}{d\theta^{'2}} \Psi(\theta') \right]^{\delta} d\theta',
$$
where $L$ is the lower bound for the parameter $\theta$ (in our case, $L=-\infty$ and $\delta$ is a constant. It is clear that if $\delta=\frac{1}{2}$, $B_{\frac{1}{2}}(\theta) = A(\theta)$ for the Rasch model. The $\delta=\frac{1}{2}$ is actually the stability of variance, which is also what we have illustrated. Other values for $\delta$ are possible and depending on those values, the transformed parameters have different properties (e.g., $\delta = \frac{1}{3}$ results in a normal likelihood). \demo

\paragraph{Miscellaneous}
In this section, we will briefly summarize two important results regarding the Riemannian metric defined on the statistical model manifold. 

\subparagraph{Fisher scoring} Let us consider the numerical problem of finding parameter estimates. Assume we have a statistical model $p(y|\theta)$ and we have collected $n$ i.i.d. samples $y_1, \dots, y_n$. If we want maximum likelihood estimates, we can define a loglikelihood $\ell(\theta) = \sum_i \log p(y_i|\theta)$ and our goal is to find: $\hat{\theta} = \arg \max_{\theta} \ell{\theta}$. If no explicit solution is available, iterative optimization routines need to be used. The most well-known are gradient (or steepest) ascent\footnote{More commonly this method is called gradient or steepest descent when one wants to minimize a function.}, Newton-Raphson and Fisher scoring. For unconstrained problems, the basic iterative step $t$ can be written as:
$$
\theta^{t+1} = \theta^t + \alpha_t C(\theta^t) \nabla \ell(\theta^t),
$$
where $\nabla \ell(\theta^t)=\left. \frac{\partial \ell}{\partial \theta} \right|_{\theta=\theta^t}$ is the gradient (also called score function) and $\alpha_t$ is a step size constant and the $C$ matrix is defined as follows:
$$
C(\theta)=
\begin{cases}
I_k & \mbox{gradient ascent}\\
H(\theta)^{-1} = \left( -\frac{\partial^2 \ell}{\partial\theta_i\partial\theta_j} \right)^{-1} & \mbox{Newton-Raphson} \\
g(\theta)^{-1} & \mbox{Fisher scoring},
\end{cases}
$$
where $I_k$ is the $k\times k$ identity matrix and $H(\theta)$ is the Hessian (or observed information) matrix.

From multivariate calculus it is well-known that the gradient $\nabla \ell(\theta)$ is perpendicular to the contour line of $\ell(\theta)$ at that point and thus the direction of the gradient is the direction of steepest ascent. Convergence based on gradient ascent can be slow when $\ell$ has ridge-like features and in such cases Newton-Raphson shows faster convergence (at the expense of being less robust). Newton-Raphson can be derived by considering a quadratic approximation to $\ell$ (hence the appearance of the Hessian matrix) and optimizing this local quadratic in each step of the algorithm. 

How does Fisher scoring fit into this picture? Why would the expected information matrix (because $g(\theta) = E(H(\theta))$) lead to a good algorithm? As is shown by \citeA{amari1998natural, amari2016information}, the step in the Fisher scoring algorithm $g(\theta)^{-1}\nabla \ell(\theta)$ is in fact in the direction of steepest ascent on the manifold. When carrying out the minimization, we are thinking of the loglikehood $\ell$ as function defined on the Euclidean parameter space $\Omega$, but to understand the properties of Fisher scoring, we need to adjust this perspective and think of the loglikelihood $\ell$ as a function defined on the model manifold $\mathcal{M}$. 

An analogy may help to make this more clear. Assume we have to find the location on the earth with the maximum temperature. For simplicity assume that the temperature varies smoothly and that there is one global maximum temperature. In this example, the earth is manifold $\mathcal{M}$ and the temperature function is $\ell$. In order to iteratively go to the location of maximum temperature, we will use a map of the earth (i.e., the parameter space $\Omega$). Although we can represent the temperature function on the map, it does not need to be true that the gradient direction on the map is the gradient direction on the manifold (only at those points where the metric becomes Euclidean).

\subparagraph{Path sampling}

A well-known problem in Bayesian statistics is to estimate the ratio of two normalizing constants. A distribution $p(y|\theta)$ can be written as: $p(y|\theta)=\frac{q(y|\theta)}{\mathcal{Z}}$, where $q(y|\theta)$ is the so-called unnormalized density (it does not integrate to 1, but to $\mathcal{Z}$, the normalizing constant). In many Bayesian problems, one can generate random draws from the distribution but has only access to $q(y|\theta)$. 

The estimation of the ratio of normalizing constants of two densities using only samples and without access to $p(y|\theta)$ is a hard problem. Let us denote the two unnormalized densities by $q_0(y|\theta)$ and $q_1(y|\theta)$ and the normalizing constants by $\mathcal{Z}_0$ and $\mathcal{Z}_1$, the goal is then to estimate:
$$
r = \frac{\mathcal{Z}_1}{\mathcal{Z}_0},
$$
based on $n_0$ and $n_1$ samples from each distribution leading to an estimator $\hat{r}$.

An identity that is of use here is the following:
$$
r = \frac{\mathcal{Z}_1}{\mathcal{Z}_0} = \frac{E_0 \left[ q_1(y|\theta) \alpha(\theta) \right]}{E_1 \left[ q_0(y|\theta) \alpha(\theta) \right]},
$$
where $\alpha(\theta)$ is an auxiliary function to be chosen such that $\mbox{var}(\hat{r})$ is minimal. The major computational difficulty in this ratio formula is that samples from one distribution will be evaluated by the non-normalized density of the other distribution. If the distributions are far apart then, this will lead to computational instabilities. Choosing the function $\alpha(\theta)$ cleverly should accommodate these problems. It turns that one can define a "bridge" distribution $q_{\frac{1}{2}}(y|\theta)$  that lies in between $q_0(y|\theta)$ and $q_1(y|\theta)$ so that $\alpha(\theta) = \frac{q_{\frac{1}{2}}(y|\theta)}{q_0(y|\theta)q_1(y|\theta)}$. As the term indicates, the idea of the bridge distribution is to make a bridge between the two distributions. In fact, multiple bridges can be considered. 

And this led \citeA{GelmanMeng1998} to the question whether a continuous path (an infinite number of bridging distributions) can be considered. The answer to this question is affirmative. In fact, the optimal path connecting the two distributions is the geodesic path we have discussed before.

\section{Curvature}

In the previous section, I  have discussed how distances can be computed on the statistical manifold. The shortest path connecting two normal distributions is generally not a straight line in the parameter space, as we are used to from Euclidean geometry. The culprit for this counter-intuitive result is that the space of probability distributions is usually curved. Briefly said, the curvature of models has to do with their intrinsic nonlinearity. However, before considering the curvature of statistical manifolds in more detail, we will first introduce some general ideas about curvature.

\subsection{Some general facts about curvature}

Although the concept curvature is easy to grasp intuitively when relating it to our everyday experience, its treatment within non-Euclidean geometry is less trivial, despite it being a key concept. The most straightforward setting to think about curvature is the case of two-dimensional surface embedded in the three-dimensional Euclidean space $E^3$ (e.g., a two-dimensional graph of the function $f(x,y)$ with $x$ and $y$ being Cartesian coordinates). For this situation, one may imagine at each point of the surface a unit normal vector perpendicular to the surface and pointing outward\footnote{The terms "outward" or "inward" have usually no meaning. But it is important that the normal vectors point consistently in a direction that leads to a differentiable vector field. Here we have chosen to denote this direction as the outward direction.}. Given a smooth surface (e.g., no sharp corners or self-intersections), this collection of normal vectors is a smoothly varying vector field. The change in the direction of the unit normal at a given point $p$ of the surface in a certain tangent direction $v$ informs us about the curvature in the direction $v$, denoted $\kappa_v(p)$. The actual computations to obtain $\kappa_v(p)$ will involve the second derivatives at the point $p$.  If $\kappa_v(p)>0$ (vs. $\kappa_v(p)<0$), then the curvature is positive (vs. negative) and if $\kappa_v(p)=0$, the surface is flat in that direction. For such a two-dimensional surface there are two important particular directions, $v_1$ and $v_2$, corresponding with the maximal and minimal curvature: $\kappa_1(p)$ and $\kappa_2(p)$.  

\begin{figure}
    \centering
\picdims[width=\linewidth]{8cm}{6cm}{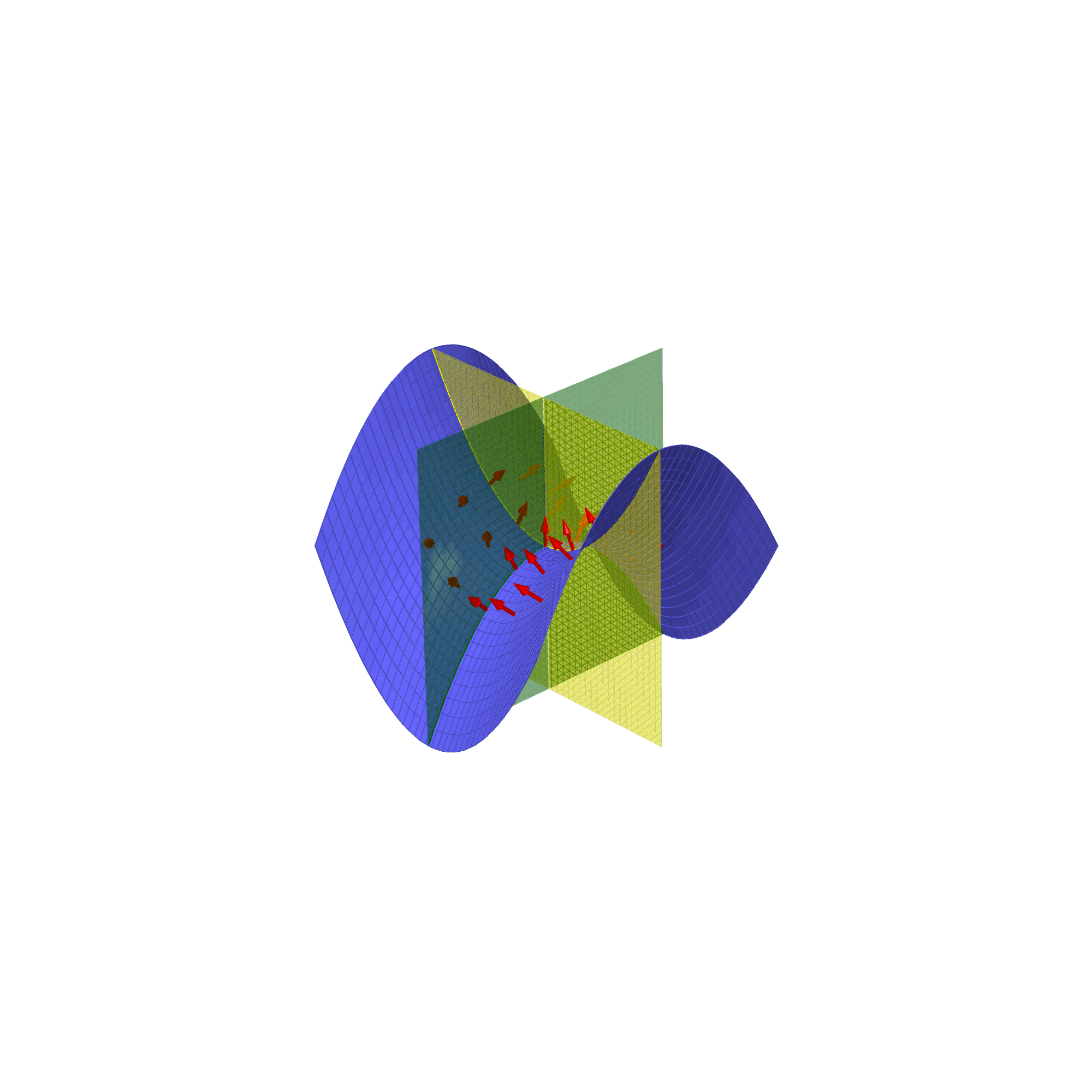}
    \caption{The graph of the function $f(x,y)=x^2-y^2$ in $E^3$ is shown together with a few selected unit normal vectors in the neighbourhood of $p=(0,0)$. The surface is also intersected by two planes ($x=0$ and $y=0$) delineating the two principal curvature directions at the intersection. The green plane defines the minimal curvature direction $\kappa_1(p)=-2$ and the yellow plane the maximal curvature direction $\kappa_2(p)=2$. The mean curvature is 0 and the Gauss curvature is $G(p)=-4$.}
    \label{fig:curvature}
\end{figure}

A sphere with radius $r$ in three dimensions has constant positive curvature: $\kappa_1(p)=\kappa_2(p)=\kappa=\frac{1}{r}>0$ for all $p$. It also shows that the larger the radius, the more the sphere locally will tend to look like a flat space (with zero curvature). A cylinder with radius $r$ in $E^3$ has $\kappa_1(p)=0$ and $\kappa_2(p)=\frac{1}{r}$ (for all $p$).  

The mean of $\kappa_1(p)$ and $\kappa_2(p)$ is the mean curvature and their product is the Gauss curvature $G(p)=\kappa_1(p)\kappa_2(p)$. Of these two curvature indices, the Gauss curvature is of particular interest. The Gauss curvature of a sphere is $G=\frac{1}{r^2}$, but for the cylinder it is 0 (e.g., equal to the Gauss curvature of a flat surface). A remarkable fact about the Gauss curvature is that it is an intrinsic property of the surface that can be calculated directly from the metric tensor $g$. This also means that any distance-preserving transformations (i.e., an isometry) does not affect the Gauss curvature. A sheet of paper can be rolled into a cylinder (keeping the Gauss curvature 0) while you cannot fold such a sheet of paper into a sphere without stretching or shrinking it (e.g., deforming the distances).

Roughly speaking, the study of curvature has been extended in two ways in differential geometry \cite{morgan2009riemannian}. A first extension is to consider $m$-dimensional surfaces embedded in $E^n$ (in which case there is not a single but multiple unit normals at each point $p$). A second extension is to remove the ambient Euclidean space. In the latter case, the goal is to study the intrinsic geometry of the space without embedding it in a higher dimensional (Euclidean) space. For example, the curvature of the surface in Figure~\ref{fig:curvature} can also be experienced and studied from the perspective of a bug living on that surface without being aware of the ambient space (the only exception are one-dimensional curves whose intrinsic curvature is always flat).

\subsection{The curvature of statistical manifolds}

To discuss the curvature of statistical models, we will follow the scheme by \citeA{transtrum2011} and start with an intrinsic account of the curvature of statistical manifolds and then consider statistical and parameter-effects curvature for curved exponential family (CEF) models. 

\paragraph{Intrinsic curvature of statistical manifolds}
Obtaining results on the intrinsic curvature requires a considerable development of technical concepts, which I will not attempt here \cite{do1992riemannian,morgan2009riemannian,needham2021visual}. Moreover, there are various ways of introducing the concept of curvature (e.g., Riemann curvature tensor, sectional curvature, Ricci curvature, scalar curvature). To keep some focus, I only mention the scalar curvature $R(p)$, which is a scalar that can be calculated for each point $p$ of the manifold. In fact, for two-dimensional surfaces (as discussed earlier), the scalar curvature is directly related to the Gauss curvature: $R(p) = 2G(p)$. Hence, if $R(p)>0$ (vs. $R(p)<0$), the manifold is positively (vs. negatively) curved at $p$ and if $R(p)=0$, then the manifold is locally flat. 

Computation of the scalar curvature\footnote{The computation starts from the metric (a $k \times k$ matrix) and requires the calculation of Levi-Civita connection coefficients ($k^3$ coefficients), the Riemann curvature tensor (of dimension $k^4$) and the Ricci curvature (dimension $k^2$).} is a tedious task, but specialized software is helpful here (for this paper we used \citeNP{sagemath}). 

\subparagraph{Single parameter distributions} Single parameter distributions always have $R=0$. As mentioned already, a bug living on a one-dimensional manifold (i.e., a curve) cannot experience any curvature. Consequently, the product of many such single parameter distributions also has $R=0$.

\subparagraph{Normal distributions}  For the class of univariate normal distributions $p(y|\theta) = \frac{1}{\sqrt{2\pi \sigma^2}}e^{-\frac{1}{2}\frac{(y-\mu)^2}{\sigma^2}}$ with parameters $\theta=(\mu,\sigma)$, it can be shown that $R = -1$. Thus, the manifold of normal distributions has constant negative curvature.

\begin{figure}
    \centering
    \begin{subfigure}[t]{0.4\textwidth}
        \centering
        \picdims[width=10cm]{3cm}{6cm}{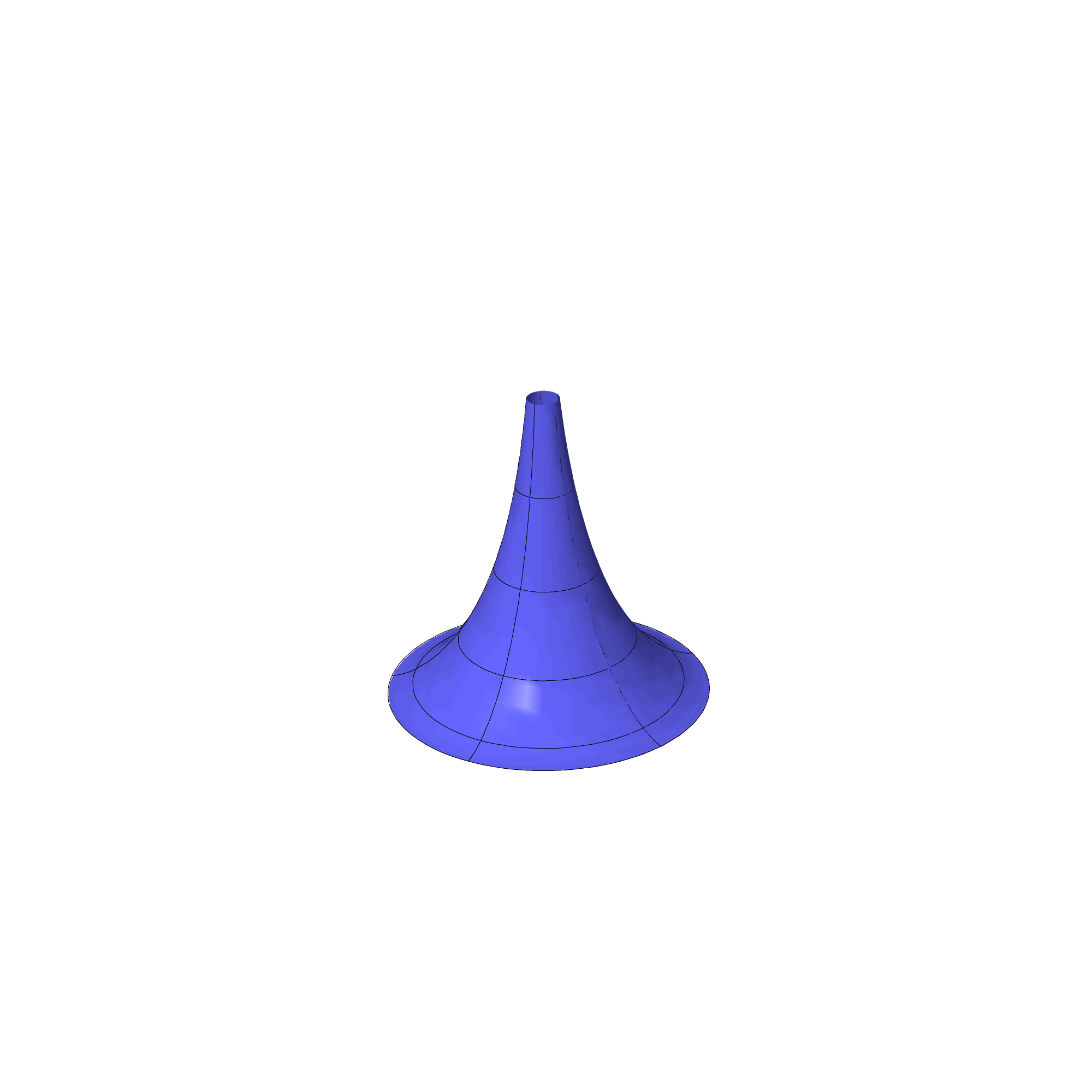}
        \caption{} \label{fig:tractritoid}
    \end{subfigure}
    \hfill
    \begin{subfigure}[t]{0.55\textwidth}
        \centering
        \picdims[width=\linewidth]{8cm}{6cm}{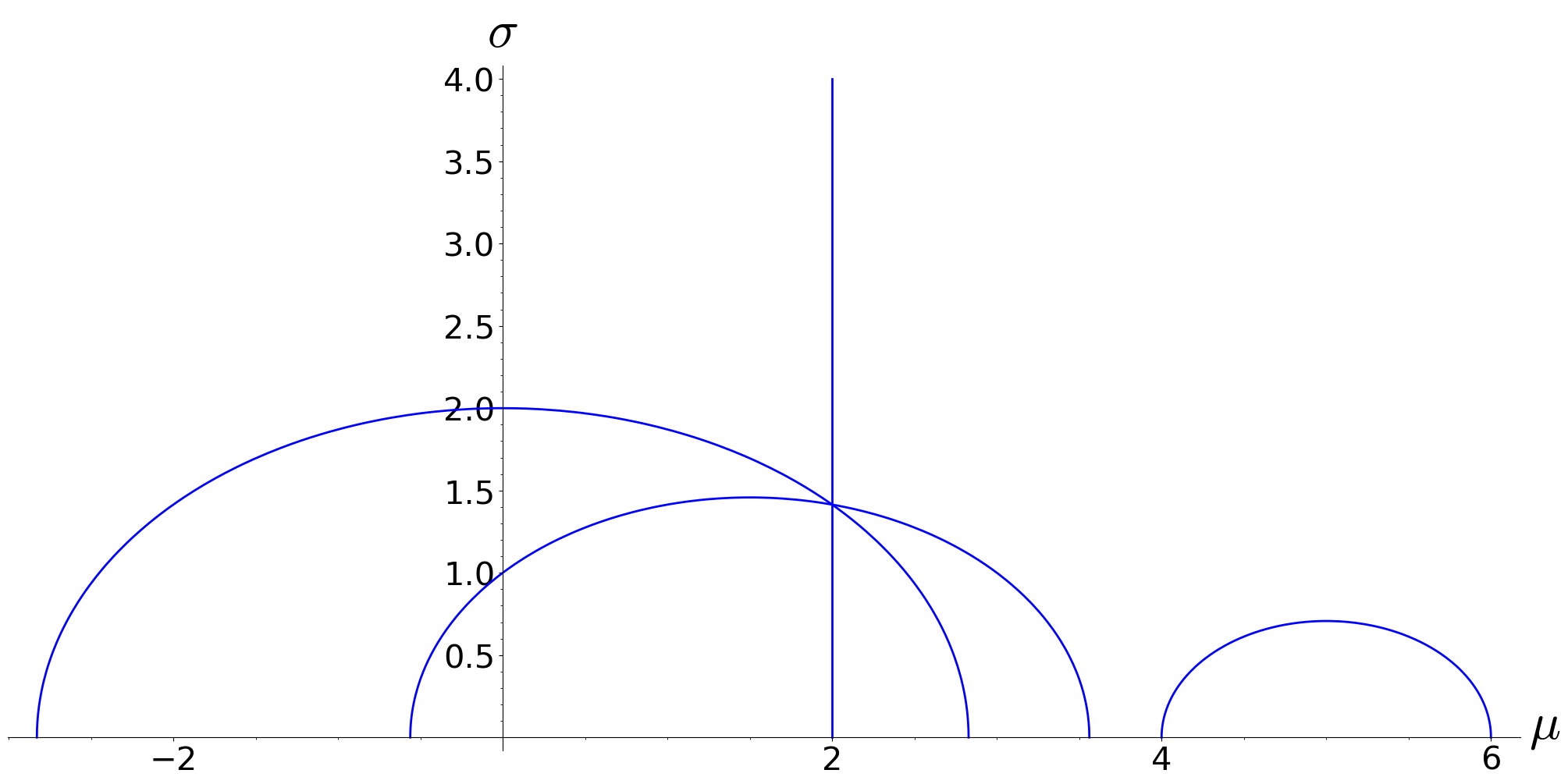}
        \caption{} \label{fig:qhyperbolic}
    \end{subfigure}
    \caption{The constant negative curvature ($R=-1$) associated with the manifold of the univariate normals $N(\mu,\sigma^2$ can be illustrated through a tractricoid (panel (a)) or the geodesics in the parameter space (panel (b)).}
    \label{fig:negcurv}
\end{figure}

It is impossible to visually represent a manifold extending infinitely in space with constant negative curvature. An imperfect physical approximating model is shown in Figure~\ref{fig:negcurv}. Panel (a) contains a tractricoid or pseudo-sphere, a surface of revolution obtained by revolving the so-called tractrix curve $(1/\cosh(t),t-\tanh(t))$ (for $t>0$)\footnote{The tractrix curve is the curve obtained by placing an object (e.g., a key) on a table, attaching a string to it and pulling the end of the string horizontally. The position of a marker on the object follows a tractrix. It is sometimes also called the "unwilling dog on the leash" curve (its German name is "Hundekurve").}. This surface has a constant negative Gauss curve of $-1$. However, this representation does not correspond perfectly the normal distribution manifold. For the tractricoid, one of the parameters (basically governing the elevation along the $z$-axis) is positive and unbounded from above (as is $\sigma$), but the other coordinate (parameter) ranges only from 0 to $2\pi$ (the revolution parameter). For the normal distribution, $\mu$ (corresponding to the horizontal dimension) ranges from $-\infty$ to $+\infty$. Thus visually representing such an unbounded surface of negative curvature is not possible.

In Figure~\ref{fig:qhyperbolic}, the $(\mu,\sigma)$ parameter space is shown together with a number of geodesic curves\footnote{The graphical representation of this half-plane is related to the famous Poincaré half-plane of hyperbolic geometry in which the geodesics are half-circles. Transforming the $\mu$ axis as follows: $\mu \rightarrow \frac{1}{\sqrt{2}}\mu$ leads to a hyperbolic space.}. As explained before, these curves are half ellipses, except for the curve connecting distributions with the same $\mu$ but different $\sigma$. The distribution corresponding to the intersection point is $N(2,(\sqrt{2})^2)$. The particular layout of curves shows that in non-Euclidean geometry, the fifth postulate of Euclid does not hold true. In a Euclidean space, given a straight line and a point (not on the line), there is a unique straight line through the point that never intersects the given line. In the non-Euclidean space in which the normal distributions live, the equivalent of straight lines are more generally the geodesic curves. The example shows that there are various geodesic curves going through the point $N(2,(\sqrt{2})^2)$ that do not intersect with the geodesic curve on the right. Hence, all three curves going through $(2,\sqrt{2})$ can be considered to be "parallel" with the rightmost curve.

Table~\ref{tab:scalarcurvnormal} contains for a number of specific normal distribution cases the scalar curvature. For the bivariate cases, our results are the same as those of \citeA{sato1979geometrical}. From this table it can be deduced that the curvature has some relation with model complexity (as measured in number of parameters) but it is not quite the same. 

\begingroup

\setlength{\tabcolsep}{10pt} 
\renewcommand{\arraystretch}{1.5} 

\begin{table}
    \centering
    \begin{tabular}{llc} \hline
      &  Distribution & Scalar curvature  \\ \hline
      1 &  Univariate normal $N(\mu,\sigma^2)$ & $-1$  \\
     2 &  Univariate normal $N(\mu,\sigma^2)$, $n$ i.i.d. observations & $-\frac{1}{n}$   \\
    3 & Bivariate normal $N(\mu,\Sigma)$  &  $-\frac{9}{2}$  \\
     4 &   Bivariate normal $N(\mu,\Sigma)$, $\rho=0$ & $-2$  \\
     5 &   Bivariate normal $N(\mu,\Sigma)$, $\sigma_1=\sigma_2, \rho=0$ & $-3$  \\
     6 &   $d$-variate normal $N(\mu,\Sigma)$, $\sigma_i=\sigma, \rho_{ij}=0$ for all $i,j$ &   $-\frac{d(d+1)}{2}$ \\ 
     7 &   $d$-variate normal $N(\mu,S)$, with $S$ known & 0 \\
      8 &  Multinomial distribution ($M=3$) & $\frac{1}{2n}$\\ 
      9 &  Multinomial distribution ($M=4$)& $\frac{3}{2n}$\\
      10 &  Multinomial distribution ($M=5$)& $\frac{3}{n}$\\ \hline
    \end{tabular}
    \caption{The scalar curvature for various distributions with constant scalar curvature.}
    \label{tab:scalarcurvnormal}
\end{table}

\endgroup
\subparagraph{Normal distributions with known $\Sigma= S$} For the class of normal distributions with a known covariance matrix (row 7 in Table~\ref{tab:scalarcurvnormal}), $p(y|\theta) = (2\pi)^{-\frac{d}{2}}(\det S)^{-\frac{1}{2}}e^{-\frac{1}{2}(y-\mu)^T S^{-1} (y-\mu)}$ and parameters $\theta=(\mu_1,\mu_2,\dots,\mu_d)$, $R=0$. Hence, in agreement with our intuition, this is completely flat manifold.

\begin{figure}
    \centering
    \begin{subfigure}[t]{0.47\textwidth}
        \centering
        \includegraphics[trim={4cm 4cm 4cm 4cm},clip,width=\linewidth]{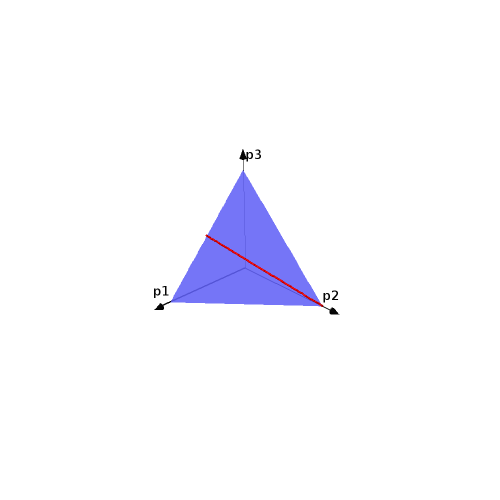}
        \caption{} \label{fig:simplex}
    \end{subfigure}
    \hfill
    \begin{subfigure}[t]{0.47\textwidth}
        \centering
        \includegraphics[trim={4cm 4cm 4cm 4cm},clip,width=\linewidth]{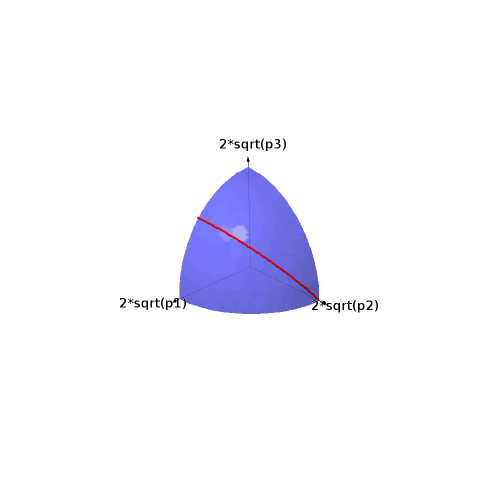}
        \caption{} \label{fig:simplexsphere}
    \end{subfigure}
    \caption{Two representations of the trinomial model with a geodesic curve (in red) between $\pi=(0.5,0,0.5)$ and $\pi=(0,1,0)$. Panel (a) shows the probability simplex and panel (b) the corresponding positive octant part of the sphere with radius 2 (after the $2\sqrt{\pi_i}$ transformation). The geodesic on the sphere is part of a great circle (i.e., it is a curve lying in the plane through the two aforementioned points and the origin.}
    \label{fig:multinom}
\end{figure}

\subparagraph{Multinomial distribution} Rows 8-10 of Table~\ref{tab:scalarcurvnormal} contain the results for the multinomial distributions. They all constitute a manifold of constant positive curvature. To explain this further, let us start with a trinomial distribution $p(y|\theta)={n \choose n_1,n_2,n_3} \pi_1^{n_1}\pi_2^{n_2}(1-\pi_1-\pi_2)^{n_3}$ with parameters $\theta=(\pi_1,\pi_2)$ leads to $R=\frac{1}{2n}$. Graphically, this model can be represented as a probability simplex (see Figure~\ref{fig:simplex}). The scalar curvature is constant and making use of the fact that for a two-dimensional manifold the scalar curvature equals twice the Gauss curvature, we arrive at the conclusion that we are dealing with a constant Gauss curvature of $\frac{1}{4n}$. A surface in $E^3$ with this property is the sphere of radius $2\sqrt{n}$, as is shown in Figure~\ref{fig:simplexsphere} (obviously, the sphere is restricted to the positive octant because of the natural bounds on $\pi$). For the general multinomial with $M$ categories, the same reasoning holds: The probability simplex is isometric with an $(M-1)$-sphere. The geodesics on the sphere are great circles. 

The intrinsic curvature of statistical manifolds as measured by the scalar curvature did not seem to have for a long time any applications in statistics, but this has recently changed. Scalar curvature shows up as a higher order term in model selection indices. Many model selection indices (e.g., AIC, BIC, stochastic complexity) have the same generic structure: A badness of fit term (e.g., $-2\log p(y|\hat{\theta})$) and terms penalizing model complexity. The first and foremost measure of model complexity is the number of parameters, but more penalty terms can be added (see e.g., \cite{Balasubramanian1997,myung2000,Balasubramanian2005}). Some of these more refined terms are rooted in the geometrical framework from this paper. Recently, \citeA{mera2022model} have shown that the scalar curvature (evaluated at the MLE) is such a higher order term that can be added to the stochastic complexity index. 

\paragraph{Statistical curvature}

In the case of extrinsic curvature (as opposed to the intrinsic curvature), we consider the curvature of a model that is a submodel of a broader class. In particular, we focus on the curved exponential family (CEF) models \cite{Efron_2023}. The multiparameter full exponential family has the following form (using the symbol $p_F$ for the density):
\begin{equation} \label{eq:fef}
p_F(y|\eta) = \exp \left( \eta^T y - \Psi(\eta) \right) p_0(y),
\end{equation}
$y$ is the $k$-dimensional sufficient statistic, $\eta \in A \subseteq \mathcal{R}^k$ the $k$-dimensional natural parameter vector and $\Psi(y)$ the multivariate log-partition function and $p_0(y)$ the carrying density. Such a specific exponential family constitutes a manifold $\mathcal{M}$ of distributions. For an exponential family model, the information matrix (or metric) can be found as follows: $g_{mn}(\eta)  =\frac{\partial^2 \Psi(\eta)}{\partial \eta_m \eta_n}$ (for $m,n = 1,\dots, k$) (see \citeNP{Efron_2023,kass2011geometrical}).

The curved exponential family model is defined by restricting the natural parameter vector $\eta$ to lie in a $q$-dimensional (with $q < k$) subspace $A_0 \subset \mathcal{R}^k$, defined by the mapping $\theta \rightarrow \eta(\theta)$, with $\theta \in \Theta \subset \mathcal{R}^q$:
\begin{equation}
    p(y|\theta) = \exp \left( \eta(\theta)^T y - \Psi(\eta(\theta)) \right) p_0(y),
\end{equation}
It is assumed that this mapping is smooth. The set of distributions defined by the CEF is denoted as $\mathcal{N}$. Some additional regularity conditions are required (see \citeNP{kass2011geometrical}), in which case $\mathcal{N}$ is a manifold embedded in the manifold $\mathcal{M}$ (i.e., $\mathcal{N} \subset \mathcal{M}$).

Several well-known statistical models are CEF models: generalized linear models, the AR(1) time series model, the mediation model, the confirmatory factor analysis model, etc.

In what follows, I will study the extrinsic curvature of the CEF in two ways. First, some analytical results on curvature for specific models will be derived. Second, the statistical curvature will be computed numerically.

\paragraph{Analytical results} For a CEF model $p_F(y|\theta)$, $\eta(\theta)$ defines a $q$-dimensional submanifold within the $k$-dimensional manifold $\eta$. The CEF model $\mathcal{N}$ is embedded in $\mathcal{M}$. The ambient space of the $q$-dimensional submanifold $\eta(\theta)$ will generally not be Euclidean, but has a metric that differs from the identity matrix (i.e., the metric of a Euclidean space).

The statistical curvature $\gamma^2_{\theta}$ is a scalar that expresses how curved the CEF is at the point corresponding with parameter value $\theta$. Originally, the statistical curvature was defined for single parameter CEF models \cite{efron1975defining}, but later it has been extended to the multivariate case by \citeA{amari1982differential}. I will directly present the formulas that can also be used in the multivariate case because most psychometric models have more than one parameter. The necessary steps to find an expression for $\gamma^2_{\theta}$ can be found in Table~\ref{tab:gamma2}\footnote{The calculation of $\gamma^2_{\theta}$ does not make use of the Levi-Civita connection (that respects lengths and angles when going from one tangent space to another) but of the more generally defined $\alpha$-connections (with $\alpha=0$ being the Levi-Civita connection, which is not used here, but instead $\alpha=1$, also called the exponential connection, is used). In fact, the geodesics of the exponential family are straight lines under this 1-connection. This is the only place in the paper where we do not use the Levi-Civita connection and hence do not respect distances and angles when going from one tangent space to another. For more information on the $\alpha$-connections, we refer to \citeA{amari2016information,calin2014geometric}.}. The calculations are also valid for single parameter models. As can be seen from Table~\ref{tab:gamma2}, first the normal vectors $u_r$ to the surface $\eta(\theta)$ are calculated (step 4) and then the inner product (with respect to the metric) of the normal vectors and the second derivatives are computed.

The most important property of $\gamma^2_{\theta}$ is that it is a measure of information loss. Focusing on maximum likelihood estimation, the MLE $\hat{\theta}$ can be seen as a statistic $S$ applied to an i.i.d. sample (of size $n$). If $\theta$ is a scalar parameter, then the relative information loss is $g^{-1}(\theta) \left[ n g(\theta) -g^S(\theta)\right]$, where $g^S(\theta)$ is the information matrix corresponding to the statistic $S$. This relative information loss tells us how much information from the complete sample is lost by compressing it into the statistic $S$. Another way of expressing this is that statistical curvature measures how much the MLE deviates from a sufficient statistic. The major result derived by \citeA{efron1975defining} is that for continuous distributions, it holds that:
$$ \label{eq:infoloss}
\gamma_{\theta}^2 = \lim_{n \rightarrow \infty} g^{-1}(\theta) \left[ n g(\theta) -g^S(\theta)\right] .
$$
The result from Equation~\ref{eq:infoloss} is only valid for the MLE. For other estimators, there is an additional term in the expression, which is not discussed here. Unfortunately, an analogue formal result for discrete distributions is not available but the statistical curvature can nevertheless be calculated. Statistical curvature can also be calculated for general distributions (i.e., not only CEF models), but we do not consider these equations here.

As a rule of thumb, \citeA{efron1975defining} argued that $\gamma^2_{\theta} > 0.125 $ can be considered high. In the same spirit, a sample size $n>8\gamma^2_{\theta}$ should wash out negative effects of statistical curvature. Results for statistical curvature for a range of models can be found in \citeA{efron1975defining}, \citeA{kass2011geometrical}, \citeA{vangarderen1999}, and \citeA{amari1982differential}. 

\begin{table}
    \centering
    \begin{tabular}{ccc} \hline
        Step & What to compute? & Explanation\\ \hline
        1 & $\dot{\eta}(\theta)_a = \frac{\partial \eta(\theta)}{\partial \theta_a}$ for $a=1,\dots,q$ & $q$ tangent vectors that span $T_{\eta(\theta)}\mathcal{N}$\\
        2 & $g_{mn}(\eta)  =\frac{\partial^2 \Psi(\eta)}{\partial \eta_m \partial \eta_n}$ for $m,n = 1,\dots, k$ & metric of the exponential model \\
        3 & $g_{ab} = \dot{\eta}_a^T g(\eta(\theta)) \dot{\eta}_b$  (for $a,b=1,\dots,q$) & induced metric for the \\
         &  & curved exponential model \\
        4 & find $u_{r}$ with $r=1,\dots,k-q$   & using Gram-Schmidt procedure \\
         & orthogonal to $T_{\eta(\theta)}\mathcal{M}$ w.r.t. the metric $g(\eta(\theta))$ & \\
        5 & $g^*_{rs} = u_r^T g(\eta(\theta)) u_s$ & induced metric on the orthogonal \\
         &  & complement of $T_{\eta(\theta)}\mathcal{M}$ \\
        6 & $H_{abr} = \frac{\partial \dot{\eta}(\theta)_a}{\partial\theta_b} g(\eta(\theta)) u_r$ & \\
        & for $a,b=1,\dots,q$ and $r=1,\dots,k-q$ & \\
        7 & $\gamma^2_{\theta} = \sum_{a,b,c,d,r,s} H_{abr} H_{cds} g_{ac}^{-1}(\theta)g_{bd}^{-1}(\theta)g^{*-1}_{rs}(\theta)$ & statistical curvature \\ \hline
    \end{tabular}
    \caption{Steps required to calculate $\gamma^2_{\theta}$ for a curved exponential family model.}
    \label{tab:gamma2}
\end{table}

\subparagraph{Example: Statistical curvature of the confirmatory factor analysis (CFA) model} As a novel application, I will compute the statistical curvature of the CFA model in a simple situation. The reason to study a simple situation is because the analytical expression of the statistical curvature becomes quickly very long. Consider three items (which are continuous random variables) with zero means. The data are $n$ three-component random vector $y_i=(y_{i1},y_{i2},y_{i3})$, i.i.d. distributed. The corresponding full exponential model in this case is a trivariate normal with zero mean vector and unstructured $3\times 3$ covariance matrix $\Sigma$:
$$
p_F(y|\Sigma) = (2\pi)^{-\frac{3n}{2}} \det(\Sigma)^{-\frac{n}{2}} \exp \left( -\frac{1}{2} \sum_i y_i^T \Sigma^{-1} y_i  \right).
$$
This is a full exponential family model. Define the precision matrix $\Phi(\eta)=\Sigma^{-1}$ as follows:
$$
\Phi(\eta) = \begin{pmatrix}
\eta_1 & \eta_2 & \eta_3\\
\eta_2 & \eta_4 & \eta_5 \\
\eta_3 & \eta_5 & \eta_6
\end{pmatrix},
$$
where $\eta=\mbox{vec}^*(\Phi(\eta))=(\eta_1,\dots,\eta_6)^T$ are its unique elements (the $\mbox{vec}^*$-operator here selects the unique elements in a rowwise manner). Then write the trivariate density function for $y$ in terms of the natural parameters $\eta$:
$$
p_F(y|\eta) = \exp \left( \eta^T t(y) + \frac{n}{2} \log \det \Phi(\eta) \right) (2\pi)^{-\frac{3n}{2}},
$$
with $t(y) = \left( -\frac{1}{2} \sum_i y_{i1}^2, -\sum_i y_{i1}y_{i2},-\sum_i y_{i1}y_{i3}, -\frac{1}{2} \sum_i y_{i2}^2,-\sum_i y_{i2}y_{i3},-\frac{1}{2} \sum_i y_{i3}^2\right)$ being the vector of sufficient statistics. The log-partition function can be written as:
$$
\Psi(\eta)= -\frac{n}{2} \log \det \Phi(\eta) =-\frac{n}{2}\log \left[\eta_1\eta_4\eta_6+\eta_2\eta_3\eta_5+\eta_2\eta_3\eta_5 - \left( \eta_3^2\eta_4+\eta_1\eta_5^2+\eta_2^2\eta_6 \right) \right].
$$

Assume now that there is a single (standardized) factor and the three items have the following loadings: 1, $\lambda$, and $\tau$. Furthermore, the error variance is $\sigma^2$ for all three items. Thus, $\theta=(\lambda,\tau,\sigma)$ and the model-implied covariance matrix is:
$$
\Sigma(\theta) = \begin{pmatrix}
1+\sigma^2 & \lambda & \tau\\
\lambda & \lambda^2+\sigma^2 & \lambda\tau \\
\tau & \lambda\tau & \tau^2+\sigma^2
\end{pmatrix},
$$
and the model-implied precision matrix is (where I have clarified that $\Phi$ depends on $\theta$ through $\eta$):
$$
\Phi(\eta(\theta)) = \frac{1}{\sigma^4 + \sigma^2(\lambda^2+\tau^2+1)} \begin{pmatrix}
\lambda^2+\tau^2+\sigma^2 & -\lambda & -\tau\\
-\lambda & 1+\tau^2+\sigma^2 & -\lambda\tau \\
-\tau & -\lambda\tau & 1+\lambda^2+\sigma^2
\end{pmatrix}.
$$
Thus, the CFA model is a CEF model with $\eta(\theta)=\mbox{vec}^*\left(\Phi(\eta(\theta))\right)$, embedded in a full exponential model. Hence, $\eta(\theta)$ (also called the solution locus by \citeNP{bates1980relative}) is a three-dimensional subspace of the six-dimensional $\eta$-space (stated differently, it is three-dimensional surface parametrized by the three-dimensional $\theta$). The embedding is nonlinear because the elements of $\eta(\theta)$ are nonlinear functions of $\eta(\theta)$.

The steps from Table~\ref{tab:gamma2} can now be followed. I will not present intermediate results as they can be obtained by hand (after some tedious algebra) or by symbolic mathematics software. The final result for the statistical curvature for the three-item CFA is then:
\begin{equation} \label{eq:statcurvCFA}
\begin{split}
\gamma^2_{\theta}  =& 2 \Bigl[16 + 128 \tau^2 + 448 \tau^4 + 896 \tau^6 + 1120 \tau^8 + 896 \tau^{10} 
   + 448 \tau^{12} + 128 \tau^{14} + 16 \tau^{16} + 128 \lambda^2 + 896 \tau^2 \lambda^2 \\
   + & 
    2688 \tau^4 \lambda^2 + 4480 \tau^6 \lambda^2 + 4480 \tau^8 \lambda^2 + 
    2688 \tau^{10} \lambda^2 + 896 \tau^{12} \lambda^2 + 128 \tau^{14} \lambda^2 + 448 \lambda^4 + 
    2688 \tau^2 \lambda^4 \\
    +& 6720 \tau^4 \lambda^4 + 8960 \tau^6 \lambda^4 + 6720 \tau^8 \lambda^4 + 2688 \tau^{10} \lambda^4 + 448 \tau^{12} \lambda^4 + 896 \lambda^6 + 4480 \tau^2 \lambda^6 + 8960 \tau^4 \lambda^6 \\
    +& 8960 \tau^6 \lambda^6 + 
    4480 \tau^8 \lambda^6 + 896 \tau^{10} \lambda^6 + 1120 \lambda^8 + 4480 \tau^2 \lambda^8 + 
    6720 \tau^4 \lambda^8 + 4480 \tau^6 \lambda^8 + 1120 \tau^8 \lambda^8 \\
    +& 896 \lambda^{10} + 2688 \tau^2 \lambda^{10} + 2688 \tau^4 \lambda^{10} + 896 \tau^6 \lambda^{10} + 448 \lambda^{12} + 896 \tau^2 \lambda^{12} + 448 \tau^4 \lambda^{12} + 128 \lambda^{14} + 128 \tau^2 \lambda^{14} \\
    +& 16 \lambda^{16} + 32 (1 + \tau^2 + \lambda^2)^7 (3 + 2 \tau^2 + 2 \lambda^2) \sigma^2 + 
    8 (1 + \tau^2 + \lambda^2)^6 (37 + 9 \tau^4 + 42 \lambda^2 + 9 \lambda^4 + 
      6 \tau^2 (7 + 3 \lambda^2)) \sigma^4 \\
    +& 16 (1 + \tau^2 + \lambda^2)^6 
     (36 + 17 \tau^2 + 17 \lambda^2) \sigma^6 + 8 (1 + \tau^2 + \lambda^2)^4 
     (99 + 70 \tau^4 + 172 \lambda^2 + 70 \lambda^4 + 4 \tau^2 (43 + 35 \lambda^2)) \sigma^8 \\
     +& 
    4 (1 + \tau^2 + \lambda^2)^3 (202 + 8 \tau^6 + 405 \lambda^2 + 
      8 \lambda^4 (26 + \lambda^2) + 8 \tau^4 (26 + 3 \lambda^2) + 
      \tau^2 (405 + 416 \lambda^2 + 24 \lambda^4)) \sigma^{10} \\
      +& 2 (1 + \tau^2 + \lambda^2)^2 
     (296 + 48 \tau^6 + 676 \lambda^2 + 429 \lambda^4 + 48 \lambda^6 + 
      3 \tau^4 (143 + 48 \lambda^2) + 2 \tau^2 (338 + 429 \lambda^2 + 72 \lambda^4)) \sigma^{12} \\ 
      +& 
    2 (1 + \tau^2 + \lambda^2) (138 + 48 \tau^6 + 353 \lambda^2 + 266 \lambda^4 + 
      48 \lambda^6 + 2 \tau^4 (133 + 72 \lambda^2) + 
      \tau^2 (353 + 532 \lambda^2 + 144 \lambda^4)) \sigma^{14} \\
      +& 
    (63 + 32 \tau^6 + 174 \lambda^2 + 146 \lambda^4 + 32 \lambda^6 + 
      2 \tau^4 (73 + 48 \lambda^2) + 2 \tau^2 (87 + 146 \lambda^2 + 48 \lambda^4)) \sigma^{16} \Bigr] \\
      \times & \Bigl[ n (1 + \tau^2 + \lambda^2)^2 (2 (1 + \tau^2 + \lambda^2)^2 + 
     4 (1 + \tau^2 + \lambda^2)^2 \sigma^2 + (3 + 4 \tau^2 + 4 \lambda^2) \sigma^4)^3 \Bigr]^{-1}.
\end{split}  
\end{equation}
The expression is quite complicated, but a few deductions can be made easily. First, the statistical curvature decreases with $n$, such that $\lim_{n\rightarrow \infty} \gamma_{\theta}^2 = 0$. Second,  $\lim_{\sigma\rightarrow 0} \gamma_{\theta}^2 = 4n^{-1}$. Third, $\lim_{\lambda \rightarrow \infty} \gamma_{\theta}^2 = \lim_{\tau \rightarrow \infty} \gamma_{\theta}^2= 2(9\sigma^4 + 8\sigma^2 + 2)\times(n(2\sigma^2 + 1)^3)^{-1}$. The last limit results for $\sigma=1$ in $38(27n)^{-1}$. 

\begin{figure}
    \centering
    \includegraphics[width=0.3\linewidth]{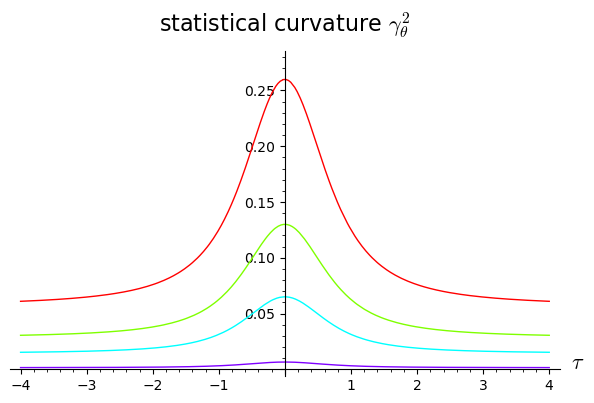} \hfill
    \includegraphics[width=0.3\linewidth]{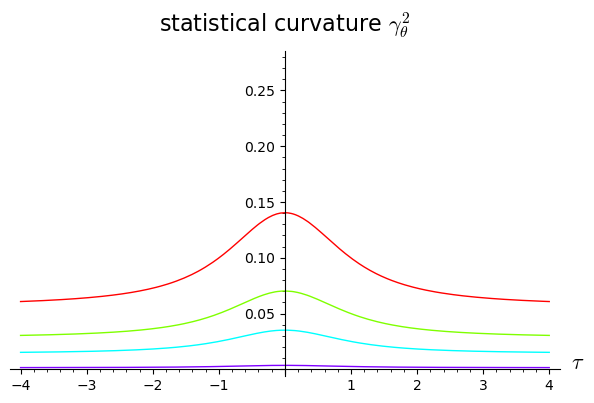}
    \includegraphics[width=0.3\linewidth]{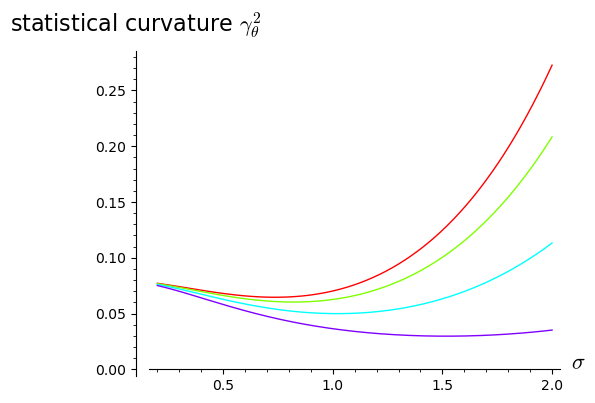}
    \caption{The left panel show statistical curvature as a function of $\tau$ ranging from $-4$ to 4, while $\lambda=0.5$ and $\sigma=1$ and the sample size $n$ is 25 (red), 50 (green), 100 (blue), and 1000 purple. The middle panel is the same, except for the value of $\lambda=1$. In the right panel $\lambda=1$ and $n=50$ while $\sigma$ varies from 0.2 to 2 and $\tau$ takes on values 0 (red), 0.5 (green), 1 (blue) and 2 (purple).}
    \label{fig:statcurvCFA}
\end{figure}

Figure~\ref{fig:statcurvCFA} displays for a number of values of $\lambda$, $\tau$ and $\sigma$ the corresponding curvature. These figures confirm the result that the theoretical results from the previous paragraph. For example, in the two left panels, the limits of the four curves are pairwise equal to $38(27n)^{-1}$ (thus, for both red curves, this equals 0.0563). The curvature becomes problematic for low sample sizes ($n$ around 25) and low loadings.

The results presented are limited to a very simple CFA model, however they carry some general value. For CFA models (and structural equation models in general), nonconvergence in small samples is a common problem \cite{dejonckererosseel2023model}. Specifically, a common algorithm is the aforementioned Fisher scoring method (which resembles Newton-Raphson except that the Fisher information matrix replaces the observed information). As shown by \citeA{smyth1987curvature} and \citeA{kass2011geometrical}, the convergence properties of the Fisher scoring method depend on the curvature: The larger the curvature, the worse the convergence. More precisely, the convergence of the sequence of iterations can be expressed as: $\lim_{k\rightarrow \infty} \frac{\theta^{k+1}-\hat{\theta}}{\theta^{k}-\hat{\theta}}=\rho$. In case $0<\rho<1$, the iterative sequence produced by the method converges to the MLE $\hat{\theta}$, while if $\rho\geq 1$, there is no convergence. \citeA{kass2011geometrical} state that $\rho \propto \sqrt{\gamma_{\theta}^2}$. In the study of \citeA{yuan2017improving}, a slightly more complex CFA model (e.g., two factors instead of one) than the one discussed above was used, but it is possible to copy their three main simulation settings: (1) $\lambda=\tau=\sigma=1$, (2) $\lambda=\tau=2$, $\sigma=1$, (3) $\lambda=\tau=1$, $\sigma=\sqrt{2}$ for $n=30$ and $n=50$. The statistical curvatures for these situations are 0.083, 0.055, 0.098 (for $n=30$) and 0.050, 0.033, 0.059 (for $n=50$). Thus based on these numbers, convergence is better (rather trivially) with larger sample size, which is the case. Less trivially, scenario (3) is the most problematic one, followed by scenario (1) and then scenario (2). These is confirmed by the simulations in \citeA{yuan2017improving} (see their Table~5, case D1). \demo

\paragraph{Numerical results} 
As can be seen from Equation~\ref{eq:statcurvCFA}, finding explicit expressions for the statistical curvature can be quite laborious and tedious. The main advantage is that it gives an expression that is not dependent on an observed data set. However, there exists also a numerical method to calculate statistical curvature (and a number of related quantities) for a model fit to a particular data set \cite{bates1980relative,kass2011geometrical}. The sequence of computation steps is given in Table~\ref{tab:numericalStatCurv}. The idea behind the computations are as follows (for more details, see \citeNP{bates1980relative,kass2011geometrical}): First, the $\eta$-space is rotated (taking into account the metric of the space) through a QR-decomposition such that the first $q$ axes span the tangent space and the remaining $k-q$ are orthogonal to it. Next, the parameters are transformed at the MLE so that the parameter grid coincides with the orthonormal basis in the tangent space. This transformation is achieved through multiplication with the $R$ matrix. That is not part of the calculations steps in Table~\ref{tab:numericalStatCurv} because ultimately, we need the derivatives with respect to the original parameters, and application of the chain rule leads to the use of the inverse matrix $L=R^{-1}$. 

Step 8 from Table~\ref{tab:numericalStatCurv} contains the formula to calculate the statistical curvature of the model at the MLE. As can be seen from the equation and the explanation, the information that is used in the equation comes from the $k-q$-dimensional orthogonal complement to the tangent space. In that sense, the equation links up nicely with the aforementioned Figure~\ref{fig:curvature}: The change in the normal (or normals if there the orthogonal subspace has dimensionality larger than one) tells something about the curvature.

The calculations outlined in Table~\ref{tab:numericalStatCurv} also result in an additional measure: $\omega_{\hat{\theta}}^2$, which is also called the parameter effects curvature. It is not an intrinsic feature of the model, but rather curvature induced by the particular choice of parameters. Across several applications, \citeA{bates1980relative} found across several data sets that $\omega_{\hat{\theta}}^2$ is larger than $\gamma_{\hat{\theta}}^2$.

Because the curvature measures $\gamma_{\hat{\theta}}^2$ and $\omega_{\hat{\theta}}^2$ 
are calculated numerically for a given data set and at the MLE, the results will vary from data set to data set. Therefore, in the application sections to follow, I will perform a couple of small simulation studies and generate $K$ (e.g., $K=1000$) data sets and summarize the calculated values $\gamma_{\hat{\theta}_j}^2$ and $\omega_{\hat{\theta}_j}^2$ (with $j=1,\dots,K$) using the harmonic mean: $\tilde{\gamma}_{\theta}^2= \frac{K}{\sum_j \left[ \gamma_{\hat{\theta}_j}^2 \right]^{-1}}$ and $\tilde{\omega}_{\theta}^2= \frac{K}{\sum_j \left[ \omega_{\hat{\theta}_j}^2 \right]^{-1}}$.

\begin{table}
    \centering
    \begin{tabular}{ccc} \hline
        Step & What to compute?  & Explanation\\ \hline
        1 & $\hat{\theta}$ & Find the MLE \\
        2 & $\dot{\eta}(\hat{\theta})  = \left( \frac{\partial \eta_m(\theta)}{\partial \theta_a}\Bigr\rvert_{\hat{\theta}} \right)=(\dot{\eta}(\hat{\theta})_a^m)$  & $k\times q$ matrix with as columns \\ 
         & for $a=1,\dots,q$ and $m=1,\dots,k$ & the $q$ tangent vectors that span $T_{\eta(\hat{\theta})}\mathcal{N}$\\
        3 & $\ddot{\eta}(\hat{\theta}) = \left(\frac{\partial^2 \eta_m(\theta)}{\partial \theta_a \partial \theta_b}\Bigr\rvert_{\hat{\theta}} \right) = \left( \ddot{\eta}(\hat{\theta})_{ab}^m \right)$  & $k\times q\times q$ array\\
         & for $a,b=1,\dots,q$ and $m=1,\dots,k$  & of second derivatives \\
        4 & $g(\eta(\hat{\theta}))=(g_{mn}(\eta(\hat{\theta}))) = h(\hat{\theta})h(\hat{\theta})^T$ & Cholesky decomposition of \\ 
        & & $k\times k$ Fisher information \\
        5 & $h(\hat{\theta})\dot{\eta}(\hat{\theta}) = (Q|N)\begin{pmatrix}
            R \\
            0
        \end{pmatrix}$ & perform QR decomposition\\
        & & $k\times q$ $Q$, $q\times q$ $R$, $k\times (k-q)$ $N$ \\
        6 & $L=R^{-1}$ & inverse of $R$ \\
        7 & $A=(A_{ab}^m)$  & $k\times q\times q$ array\\
        & $= \left( \sum_{n,o} \sum_{a',b'} Q_{nm} h(\hat{\theta})_{on} L_{aa'} L_{bb'} \ddot{\eta}(\hat{\theta})_{a'b'}^o  \right)$ & \\
        8 & $\gamma_{\hat{\theta}}^2 = \sum_{a,b} \sum_{m=q+1}^k (A_{ab}^m)^2$ & statistical curvature calculated from \\
        & & $k-q$-dimensional orthogonal complement \\
        & &  to the tangent space \\
        9 & $\omega_{\hat{\theta}}^2 = \sum_{a,b} \sum_{m=1}^q (A_{ab}^m)^2$ & parameter-effects curvature calculated from \\
        & & $q$-dimensional tangent space  \\
        \hline
    \end{tabular}
    \caption{Steps required to numerically evaluate $\gamma^2_{\hat{\theta}}$ (and the parameter-effects curvature  $\omega^2_{\hat{\theta}}$) for a curved exponential family model.}
    \label{tab:numericalStatCurv}
\end{table}

\subparagraph{Example: Numerical calculation of the statistical curvature of the confirmatory factor analysis (CFA) model} Let us continue with the simple CFA model considered earlier. Based on the similar three scenarios as considered earlier ((1) $\lambda=\tau=\sigma=1$, (2) $\lambda=\tau=2$, $\sigma=1$, (3) $\lambda=\tau=1$, $\sigma=\sqrt{2}$), $n=30$ and $K=1000$, we find that $\tilde{\gamma}_{\theta}^2$ equals 0.087, 0.058, and 0.095, respectively. For the parameter-effects curvature, $\tilde{\omega}_{\theta}^2$ equals 0.316, 0.326, and 0.345. For $n=50$, $\tilde{\gamma}_{\theta}^2$ equals 0.051, 0.034, and 0.057, respectively, while $\tilde{\omega}_{\theta}^2$ is estimated to be 0.190, 0.196, and 0.206. Comparing with the theoretical values, the estimated statistical curvatures are very similar. It can also be concluded that the parameter-effects curvature $\tilde{\omega}_{\theta}^2$ is much larger than he statistical curvature $\tilde{\gamma}_{\theta}^2$, even for small samples.\demo

\subparagraph{Example: Numerical calculation of the statistical curvature of an IRT model} As a next example, we will consider an IRT model. I have simulated a single data set (with $n=500$ test takers) from the following model:
$$ \label{eq:2PLM}
\Pr(y_{pj}=1) = \frac{e^{\alpha_{a[j]}(\theta_{g[p]}-\beta_{b[j]})}}{1+e^{\alpha_{a[j]}(\theta_{g[p]}-\beta_{b[j]})}},
$$
where $\alpha$, $\beta$, and $\theta$ are the discrimination, difficulty and ability parameter. To keep the computational burden under control, the number of parameters is strongly reduced. There are only two values for the discrimination parameter: $\alpha_1=1$ and $\alpha_2$ (with one equal to 1 for identification reasons). Also, there are only two values for the difficulty parameter: $\beta_1=0$ and $\beta_2$ (with one equal to 0 for identification reasons). There are also only two ability groups: $\theta_1$ and $\theta_2$. Each parameter value is assigned to half of the persons or items and all combinations are crossed. The functions $a[j]$, $b[j]$, and $g[p]$ are selector functions that select the appropriate parameter. As a consequence, the number of parameters is lower in this version of the model than the original one, but the nonlinearity is still present.

\begin{figure}
    \centering
    \includegraphics[width=0.75\linewidth]{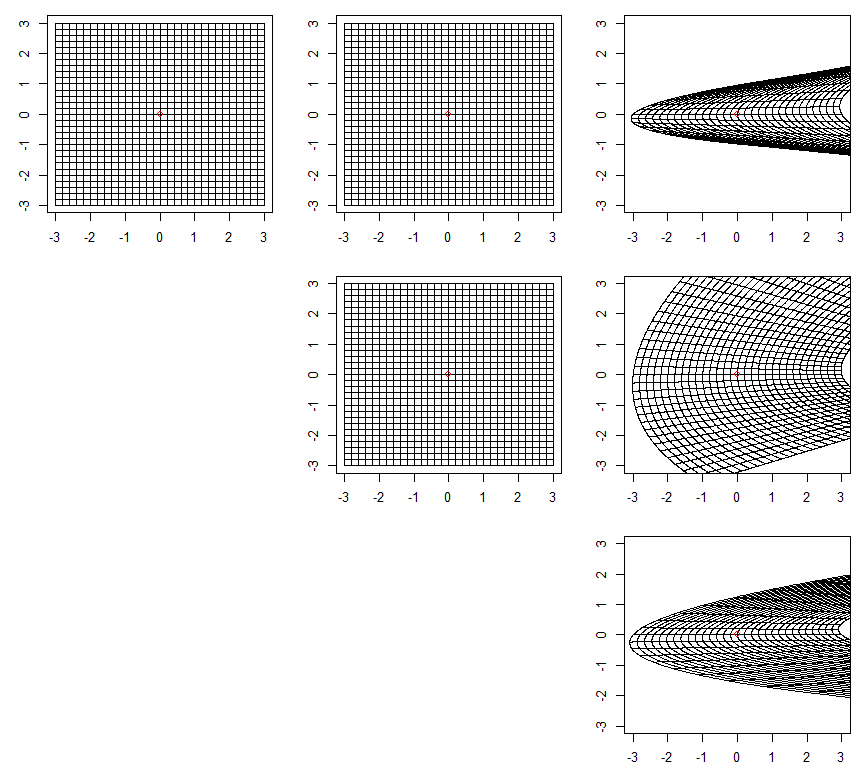}
    \caption{Illustration of parameter-effects curvature for the two-parameter logistic model.}
    \label{fig:curv2PLM}
\end{figure}

The calculations show that $\gamma^2_{\hat{\theta}}=1.02\cdot 10^{-27}$ and $\omega^2_{\hat{\theta}}=59.25$. Hence, the intrinsic statistical curvature is negligible with $n=500$, while the parameter-effects curvature is rather large. The latter is further illustrated in Figure~\ref{fig:curv2PLM} which shows the projection of the parameter curves onto the tangent space (at the MLE, indicated by red circle in the middle of the plot). Because the tangent space is four-dimensional, it cannot be visualized, but the figure contains pairwise plots.\demo

\section{Volume}

The final geometrical aspect I want to discuss is volume. Using the metric, not only distances can be calculated on the manifold but also volumes. In order to give some background, I will be using elements from \citeA{Boothby1986} and \citeA{catichaEntropic} (see also \citeNP{segert2019general} and \citeNP{Kristiadi2023}).


At a specific point (or density) $p$ on the manifold\footnote{For integration, it is required that the manifold is oriented. This is not problematic because a statistical manifold, which are fully characterized by a single parametrization, is automatically oriented.}, denoted $p(y|\theta)$ in the $\theta$ parametrization and with metric $g(\theta)$, we can now look at the volume in a small region around $p(y|\theta)$ by considering a $k$-dimensional parallelepiped on the manifold spanned by considering a small change in each of the coordinates in turn: $d\theta_i$ (for $i=1,\dots,k$). This volume element is denoted $dV_g(\theta)$. In order to find the volume of the parallelepiped, first a local change of coordinates from $\theta$ to $\xi$ is made, such that $A$ is the Jacobian matrix of the transformation from $\xi$ to $\theta$ evaluated at the current point. For simplicity, we assume that the transformation is orientation preserving, so that $\det A>0$. The $\xi$ coordinates are chosen such that this $\xi$-frame of reference is locally an orthonormal Cartesian coordinate system with metric $g(\xi)=I_k$ ($I_k$ being the identity matrix). Therefore, the parallelepiped reduces to a hypercube with side lengths 1: $dV_I(\xi) = d\xi_1 d\xi_2 \dots d\xi_k$ and hence the volume is 1. Expressing the volume of the original parallelepiped (i.e., $dV_g(\theta)$) under the change of coordinate system gives: $dV_g(\theta) = \det A d\theta_1 d\theta_2 \dots d\theta_k$.

Under the same transformation, the expression for the metric also changes locally. Using the change-of-coordinates formula for the metric (see Equation~\ref{eq:transformfisher} for the unidimensional counterpart): $g(\theta) = A^T I_k A = A^T A$. Therefore, $\det(g(\theta)) = \det A^T A=(\det A)^2$. From this it follows that $\det A = \sqrt{\det g(\theta)}$. Finally, we obtain that
\begin{equation}
dV_g(\theta) = \sqrt{\det g(\theta)} d\theta_1 d\theta_2 \dots d\theta_k,  \label{eq:volel}
\end{equation}
where $dV_g(\theta)$ is called the Riemannian volume form \footnote{Because the Fisher information matrix is positive definite, no absolute value has to be taken under the square root.}\footnote{A more detailed treatment of this subject requires the use of differential forms. We refer to \cite{lee2003introduction,oneill1997elementary,needham2021visual}.} In sum (paraphrasing \citeNP[p.224]{oneill1997elementary}), the elements of Fisher information can be considered as "warping functions" because they measure the way the flat parameter space $\Omega \subseteq \mathbb{R}^k$ gets distorted into the curved manifold $\mathcal{M}$.

With the Riemannian volume form, we can perform integration on the manifold. As an application, I will discuss Jeffreys' prior.

\paragraph{Example: Jeffreys' prior} The volume element $dV_g(\theta)$ can be considered as a measure on the manifold, leading to a distribution $p^{\mathcal{M}}(\theta)$ on the manifold. Starting with a prior $p(\theta)$ on the parameter space, the probability in a small region around $\theta$ should the same as the probability around $p(y|\theta)$ on the manifold:
$$
p^{\mathcal{M}}(\theta) dV_g(\theta) = p(\theta) d\theta_1 d\theta_2 \dots d\theta_k.
$$
Moving the volume element $dV_g(\theta)$ to the right-hand side replacing by its expression from Equation~\ref{eq:volel} then gives:
\begin{align}
p^{\mathcal{M}}(\theta) &= p(\theta)\frac{d\theta_1 d\theta_2 \dots d\theta_k}{dV_g(\theta)}  \nonumber \\
&=p(\theta)\frac{d\theta_1 d\theta_2 \dots d\theta_k}{\sqrt{\det g(\theta)} d\theta_1 d\theta_2 \dots d\theta_k} \nonumber \\
&=\frac{p(\theta)}{\sqrt{\det g(\theta)}}.  \label{eq:mfddistr}
\end{align} 
If we want the prior to assign equal weight to all distributions at the level of the manifold (i.e., $p^{\mathcal{M}}(\theta) \propto 1$) , then Equation~\ref{eq:mfddistr} tells us that we have to take $p(\theta) = \sqrt{\det g(\theta) }$, which is Jeffreys' prior. If the integral $\int_{\Omega} \sqrt{\det g(\theta) } d\theta_1 d\theta_2 \dots d\theta_k$ diverges, Jeffreys' prior is improper (which happens for non-compact manifolds).

A key feature of Jeffreys' prior is its parametrization invariance. This means that the same rule (i.e., the square root of the determinant of the Fisher information matrix) can be used in any transformation (as long as it is a smooth transformation of $\theta$). Assume that $\phi =  \phi(\theta)$, and thus $\theta = \theta(\phi)$. The Fisher information matrix expressed in the $\phi$-parameters is $g(\phi)= \left( \frac{\partial \theta}{\partial \phi} \right)^T g(\theta) \left( \frac{\partial \theta}{\partial \phi} \right) $, where $\frac{\partial \theta}{\partial \phi}$ is the Jacobian of the transformation matrix of the transformation $\theta(\phi)$. Then it follows that:
\begin{align*}
 \sqrt{\det g(\phi)} &=\sqrt{\det \left[ \left( \frac{\partial \theta}{\partial \phi} \right)^T g(\theta(\phi)) \left( \frac{\partial \theta}{\partial \phi} \right) \right] }  \\
 &=\sqrt{\left( \det \frac{\partial \theta}{\partial \phi} \right)^2 \det g(\theta(\phi))} \\
 &=\left| \det \frac{\partial \theta}{\partial \phi} \right| \sqrt{\det g(\theta(\phi))}.
\end{align*}
From this, the Jacobian determinant can be expressed as: $\left| \det \frac{\partial \theta}{\partial \phi} \right| = \frac{ \sqrt{\det g(\phi)}}{\sqrt{\det g(\theta(\phi))}}$. If we then consider the prior in the $\phi$-parametrization as a transformation from the Jeffreys' prior in the $\theta$-parametrization, we obtain:
\begin{align*}
p(\phi) &= p(\theta(\phi)) \left| \frac{\partial \theta}{\partial \phi} \right| \\
&= \sqrt{\det g(\theta(\phi))} \frac{ \sqrt{\det g(\phi)}}{\sqrt{\det g(\theta(\phi))}} \\
&=\sqrt{\det g(\phi)},
\end{align*}
which is again the same formula for Jeffreys' prior. Hence, also in the $\phi$-parametrization, the rule for the obtaining the prior remains the same and leads to a uniform prior on the manifold.

As shown by \citeA{GEORGE1993169}, Jeffreys' prior is not the only parametrization-invariant construction rule. In fact, Jeffreys' prior belongs to a wider class of invariant priors that are derived from considering a certain discrepancy between distributions. For example, Jeffreys' prior is linked to the Kullback-Leibler divergence and the Hellinger distance, but one could also use a squared Euclidean distance and this leads to another invariant prior. However, these other priors do not result in a uniform distribution on the manifold but assign the probability weights differently.  \demo

\section{Conclusion}

In this paper, I have tried to provide a geometrical perspective on parametric psychometric models by discussing three key geometrical features (distance, curvature and volume) with respect to statistical models. Many of the results presented in this paper have been derived by other researchers in the past decades, but a few applications of the methods to psychometric model are new. 

A geometric perspective puts the emphasis on aspects and properties of models that are invariant to reparametrization. The idea is similar to the situation in physics where one wants to use laws that are independent of the choice of a particular coordinate system\footnote{Specifically for general relativity, the following quote by Einstein is relevant here (because it pertains to Riemannian geometry as also used in this paper): "Now it came to me: … the independence of the gravitational acceleration from the nature of the falling substance, may be expressed as follows: In a gravitational field (of small spatial extension) things behave as they do in a space free of gravitation. … This happened in 1908. Why were another seven years required for the construction of the general theory of relativity? The main reason lies in the fact that it is not so easy to free oneself from the idea that coordinates must have an immediate metrical meaning." \cite[pp.65-67]{schilpp1959albert}.}. Obviously, to carry out computations, a coordinate system must be set up (and it is often chosen in such a way that it allows for easy computations) but the final result may not depend on this arbitrary choice. 

Thus, the mathematical function of the coordinate system in physics and the parameter space in psychometrics (or statistics for that matter) can be considered analogue: In physics, the coordinate system is used to refer or to identify points on the spacetime manifold, while in statistics, the coordinate system (i.e., parametrization) is used to identify probability distributions. However, it is clear that in both fields, the meaning of the coordinate systems is also very different. In psychometrics, the question of interest is linked to particular parameters, or the parameters have a particular meaning in terms of the underlying generative processes. But even in those situations, it is often difficult to argue for a specific choice of parameters. 

In his provocative book, \citeA{taagepera2008making} makes another comparison between parameters (or "adjustable constants" as he calls them) in statistics and physics. On the one hand, parameters in physics are not the coordinates but the fundamental constants of nature (e.g., the gravitational constant, Planck's constant, the velocity of light, etc.). The number of parameters in physics equations is rather small and their role is to glue various equations together (connecting quantities with different dimension). On the other hand, the number of parameters in the behavioral and social sciences are typically large and they most often are not used to link different equations. 

Another, and often proposed, way of making claims invariant with respect to a specific parametrization is by focusing on predictions. This has been advocated by several statisticians \cite{geisser2017predictive,billheimer2019predictive,shmueli2010explain} and it is also the core tenet of artificial intelligence. In such situations, the interpretation of parameters is of no particular interest. This is somewhat similar to the situation in artificial intelligence (AI), in which case the number of parameters in modern applications runs in the orders of million or even billions, effectively prohibiting any meaningful interpretation.

Although prediction and predictive inference tools are used in psychometrics and statistics, parameters usually do play an important role in the scientific process. Roughly speaking, there are two types of models. The first type are models with relatively few parameters (e.g., the diffusion model for speeded perceptual decision making, \citeNP{ratcliff1978theory}). Such models are closer in spirit to physics. A second type of models have a quite large number of parameters (e.g., item response models, mixed models, structural equation models, etc.) that contain a larger number of parameters (but far less than the number of parameters in overparametrized AI models). However, in both situations, parameters represent underlying psychological processes.

This focus of this paper has been on results that are invariant with respect to the chosen parametrization. Taking this point of view to the extreme, one may wonder whether all parametrizations, as long as they can be used to index the distributions on the manifold, are equal. Obviously, this is not the case. In many cases, researchers have stated their question of interest in terms of a specific parameter or set of parameters in a statistical model. In addition, manipulations are set up that target one specific parameter and not the others (e.g., \citeNP{voss2004interpreting}). This is called selective influence and if such an exercise succeeds, then it clearly adds credibility to the interpretation of and the choice for a particular parametrization. 

Even in situations with clearly preferred parametrization, there is fairly often a distinction between the key parameters of interest and those that that mainly present to obtain an adequate model fit without having a clear substantive interpretation and that that are often difficult to estimate. \citeA{transtrum2011} use the terminology "sloppy models" and "sloppy parameters". These sloppy parameters are not well-constrained by the data. 

This being said, there is much more to discover in the broader field of information geometry. Inevitably, a number of open questions remain and they can serve as signposts for future scholars in the field. With respect to distance, a new parametrization-invariant ability $A(\theta)$ has been derived. However, when considering the Rasch model, what is its relation to specific objectivity (i.e., the principle that you can compare two persons regardless of the item)? In addition, the new ability is dependent on the number of items, which raises the question how to compare abilities from tests with different lengths. With respect to (statistical) curvature, for which psychometric models and to which extent does it play a role in inference (e.g., estimation, convergence of algorithms, model selection, etc.)? Finally, volume offers the possibility to define a measure on the manifold, but how we use this measure in another way than only requiring a uniform over the manifold (i.e., can we develop tools to define other useful distributions over the manifold that carry meaningful information)?


Considering statistical models as manifolds with a structure that allows a geometrical study is a fascinating idea. However, it is a difficult subject as well and due to the limited competence of the author, the results in this paper are of similar size. This paper maybe an inspiration to other, more competent, researchers to explore this domain further.

\newpage
\

\medskip

\bibliographystyle{apacite}
\bibliography{bibinfgeom}

\end{document}